\documentclass[12pt,preprint]{aastex}

\renewcommand {\deg}   {\mbox{$^\circ$}}

\newcommand   {\kms}   {\mbox{km\,s$^{-1}$}}
\renewcommand {\ga}    {\mbox{\rlap{\hbox{\lower5pt\hbox{$\sim$}}}\hbox{$>$}}}
\renewcommand {\la}    {\mbox{\rlap{\hbox{\lower5pt\hbox{$\sim$}}}\hbox{$<$}}}

\begin{document}

%\title{ }
%\author{ }
%\email{ }
%\altaffiltext{1}{ }

%\begin{abstract}
%\vspace {5pt}
%\end{abstract}

%\keywords{ }
%\documentclass[12pt,preprint]{aastex}
%% manuscript produces a one-column, double-spaced document:
%% \documentclass[manuscript]{aastex}
%% preprint2 produces a double-column, single-spaced document:
%\documentclass[preprint2]{aastex}
%% Sometimes a paper's abstract is too long to fit on the
%% title page in preprint2 mode. When that is the case,
%% use the longabstract style option.
%% \documentclass[preprint2,longabstract]{aastex}
\def\kms {\hbox{km{\hskip0.1em}s$^{-1}$}} % km/s
\def\msol{\hbox{$\hbox{M}_\odot$}}
\def\lsol{\hbox{$\hbox{L}_\odot$}}
\def\kms{km s$^{-1}$}
\def\Blos{B$_{\rm los}$}
\def\etal   {{\it et al. }}                     % et al
\def\psec           {$.\negthinspace^{s}$}
\def\pasec          {$.\negthinspace^{\prime\prime}$}
\def\pdeg           {$.\kern-.25em ^{^\circ}$}
\def\degree{\ifmmode{^\circ} \else{$^\circ$}\fi}
\def\ee #1 {\times 10^{#1}}          % \ee p       10^p
\def\ut #1 #2 { \, \textrm{#1}^{#2}} % \ut unit p  unit^p
\def\u #1 { \, \textrm{#1}}          % \u unit     unit
\def\nH {n_\mathrm{H}}

\def\ddeg   {\hbox{$.\!\!^\circ$}}              % Degrees over dot
\def\deg    {$^{\circ}$}                        % Degrees symbol
\def\le     {$\leq$}                            % <=
\def\sec    {$^{\rm s}$}                        % Second of time
\def\msol   {\hbox{$M_\odot$}}                  % Solar mass
\def\i      {\hbox{\it I}}                      % italic I
\def\v      {\hbox{\it V}}                      % italic V
\def\dasec  {\hbox{$.\!\!^{\prime\prime}$}}     % Arcseconds over dot
\def\asec   {$^{\prime\prime}$}                 % Arcseconds symbol
\def\dasec  {\hbox{$.\!\!^{\prime\prime}$}}     % Arcseconds over dot
\def\dsec   {\hbox{$.\!\!^{\rm s}$}}            % Second over dot
\def\min    {$^{\rm m}$}                        % Minutes of time
\def\hour   {$^{\rm h}$}                        % Hours of time
\def\amin   {$^{\prime}$}                       % Arcminutes symbol
\def\lsol{\, \hbox{$\hbox{L}_\odot$}}
\def\sec    {$^{\rm s}$}                        % Second of time
\def\etal   {{\it et al. }}                     % et al.

\def\xbar   {\hbox{$\overline{\rm x}$}}         % bar over x

%\slugcomment{Submitted to ApJ}
\shorttitle{structure function}
\shortauthors{zadeh}

\title{Rapid Intrinsic  Variability of  Sgr A* at Radio Wavelengths}
\author{F. Yusef-Zadeh$^1$, M. Wardle$^2$, J. C. A.  Miller-Jones$^{3,4}$,
D. A. Roberts$^5$, N. Grosso$^6$ \& D. Porquet$^6$}
\affil{$^1$Department of  Physics and Astronomy, Northwestern University, Evanston, IL 60208}
\affil{$^2$Department of Physics ad Astronomy, Macquarie University, Sydney NSW 2109, Australia}
\affil{$^3$NRAO, Charlottesville, 520 Edgemont Road, VA 22903}
\affil{$^4$International Center for Radio Astronomy Research - Curtin University, GPO Box U1987,
Perth, WA 6845, Australia}
\affil{$^5$Adler Planetarium \& Astronomy Museum 1300 S. Lake Shore Drive,  Chicago, IL 60605}
\affil{$^6$Observatoire astronomique de Strasbourg, Universit\'e de Strasbourg, CNRS, UMR 7550,
11 rue de l'Universit\'e, F-67000 Strasbourg, France}

\begin{abstract}
Sgr A* exhibits flares in radio, millimeter and submm wavelengths with
durations of $\sim 1$ hour.  Using  structure function, power spectrum and 
autocorrelation function analysis, we
investigate the variability of Sgr A* on time scales ranging from a few
seconds to several hours  and find evidence for sub-minute time scale
variability at radio wavelengths.  These measurements suggest a strong case
for continuous variability from sub-minute to hourly time scales.  This
short time scale variability constrains the size of the emitting region to
be less than 0.1\,AU. Assuming that the minute time scale fluctuations of
the emission at 7\,mm arise through the expansion of regions of optically
thick synchrotron-emitting plasma, this suggests the presence of explosive,
energetic expansion events at speeds close to $c$.  The required rate of
mass processing and energy loss of this component are estimated to be $\ga
6\times 10^{-10} \msol$ yr$^{-1}$ and 400\,$L_\odot$ respectively.  The
inferred scale length corresponding to one-minute light travel time is
comparable to the time averaged spatially resolved 0.1AU scale observed at
1.3mm emission of Sgr A*.  This steady component from Sgr~A* is
interpreted mainly as an ensemble average of numerous weak and overlapping
flares that are detected on short time scales.  The nature of such short
time scale variable emission or quiescent variability is not understood but
could result from fluctuations in the accretion flow of Sgr~A* that feed
the base of an outflow or jet.
\end{abstract}


\section{Introduction}
\label{introduction}

The compact radio source Sgr A* is located at the very dynamical center of our galaxy 
(Reid and Brunthaler 2004) and is established to coincide with a 4 $\times 10^6$\msol\ black hole 
(Ghez et al. 2008; Gillessen et al. 2009). 
It is well 
known that the bolometric luminosity of Sgr~A* ($\sim$100 \lsol) is several orders of magnitude lower than 
expected for the estimated accretion rate.  A number of theoretical models have been proposed to explain 
its very low radiative efficiency and matching the spectral energy distribution (SED) of its quiescent 
emission (Melia \& Falcke 2001; Yuan et al. 2003; Liu et al.2004).
To address its underluminous nature, our approach has been 
to study the time variability of Sgr~A*.

The bulk of the continuum flux of Sgr~A* is believed to be generated in an accretion disk, where the 
source of variable continuum emission is also localized.  Recent MHD simulations have also indicated that 
variability is a fundamental property of emission in the disk of Sgr~A* 
(Hawley et al. 2002; Goldston et al. 2005; Chan et al. 2009; Moscibrodzka et al. 2009). 
The variable optical continuum flux of AGNs is also 
known to signal activities of the central engine and is detected throughout the entire electromagnetic 
spectrum with periods ranging from days to years (Arshakian et al. 2010). Sgr~A*, being a hundred 
times closer  than the next nearest example, provides us with the best view of the 
the accretion disk surrounding a supermassive black hole.  In addition, the relatively low mass of Sgr A* compared to those 
in AGNs presents an unparalleled opportunity to study  its temporal characteristics from minutes to years 
and investigate the process 
by which gas is captured, accreted or ejected.  As the dynamical time scales with the mass of a black hole, the time scale for 
variability can be argued to be proportional to the mass of the black hole.  Here, we study light curves 
of Sgr A* based on several days of observations at 7 and 13mm as well as 
radio, IR and X-ray data taken simultaneously on 2007, April 4. These measurements 
will  characterize the time scale of  variability at multiple wavelengths. 

\section{Analysis}

The structure function (SF) 
is defined as the mean difference of pairs of flux 
measurements 
separated by time lag $\tau$, i.e. $<[S(t) - S(t+\tau)]^2>$ 
(e.g., Simonetti et al. 1985; Hughes, Aller \& Aller 1992). 
It has been argued that the slope of the SF (where 
SF$\propto\tau^{\beta}$) is  related 
 to the index of the power spectrum of fluctuations $\alpha$ 
where power spectrum P 
$\propto f^{-{\alpha}}$ (Do et al. 2009). Breaks in the power spectra are seen in both AGN and X-ray binaries, 
with the 
break timescale 
scaling with black hole mass and bolometric luminosity (i.e. mass accretion rate; McHardy et al.(2006).  Using SF 
analysis, previous IR measurements have characterized the intrinsic variability time scale of Sgr~A* (Do et al.  2009) 
and have determined the break frequency in the power spectral density distribution or a turnover (rollover) time scale 
in the structure function (Meyer et al. 2009).
 This technique is widely used in the study of AGN light curves in X-rays 
(Kataoka et al. 2001).  However, a recent study by Emmanoulopoulos et al. 
(2010) concludes that any break timescales 
derived from SFs are doubtful as they depend very much on the length and underlying power spectral density of the data 
sets used. Thus, this paper does not investigate the nature of breaks 
in the structure function and  uses different analysis techniques to confirm the behavior of the calculated structure 
functions at short time scales.

To address the time variability of Sgr A*, we use three  different methods of analysis, namely, structure function, 
power spectra and autocorrelation function. 
Two different types of power spectrum are also calculated. 
One type of power spectrum is subjected to the CLEAN deconvolution algorithm and the 
power spectra are 
calculated using the procedure documented in Appendices B and C of Roberts, Lehar \& Dreher (1987). The CLEAN power 
spectrum is shown in equally-spaced bins in log space, with 30 bins covering the range between $\nu  = 10^{-2}$ 
min$^{-1}$ 
and 10 min$^{-1}$. Another type of power spectrum follows the prescription given by Uttley et al. (2002). This 
derivation, which does not include any CLEAN deconvolution,  
computes the power spectrum from
$\nu_{min}$ (1/T when T is the length  of the light curve in the time domain) 
up to the Nyquist frequency and 
normalizes it such 
that the integration from $\nu_1$ to $\nu_2$  gives the
contribution to the fractional rms squared variability on time scales $\nu_2^{-1}$ to $\nu_1^{-1}$.  
Both power spectra are binned and power law slopes are  fitted 
into the entire spectrum. 
In addition, we present  autocorrelation function 
which can potentially provide 
information on the nature of the physical process causing any observed variability.  
The auto-correlation analysis uses the Z-transformed discrete correlation function (ZDCF)
algorithm (Alexander 1997). A maximum  in the likelihood value is identified at a zero time lag. 
The ZDC function  is an improved solution to the problem of investigating correlation in unevenly sampled light 
curves. The 
standard solutions are interpolation of the existing light curve, which is considered to be unreliable when power 
exists on smaller timescales than the gaps and binning the data using discrete correlation functions (e.g., 
Edelson \& Krolik 1988). 
 Here, we first investigate 
the nature of time variability of Sgr A*  in radio wavelengths using these three  different statistical analysis. 
We then compare  structure function and CLEAN power spectrum  of IR,  X-ray and radio data taken 
simultaneously on  2007, April 4. 

%The normalization applied in these caclulations is  different than those of 
%Robert et al. (2007). 

\section{Radio Variability}
\subsection{Observations}

Radio observations were taken with the Very Large Array (VLA)\footnote{The
National Radio Astronomy Observatory is a facility of the National Science
Foundation operated under cooperative agreement by Associated Universities,
Inc.} in its C configuration on 2008 May 5-6, 10-11.  We also analyzed
archival data taken with the VLA in its A configuration on 2006
February 10 at 7mm.  Briefly, we used a fast-switching technique to observe
Sgr A* simultaneously using three sub-arrays in the C configuration at 7,
13 and 35mm.  We cycled between the fast-switching calibrator 17443-31166
(2.3 degrees away from Sgr A*), 17456-29004 (Sgr A*) for 30s and 90s,
respectively, throughout the observation.  In addition, we observed the
phase calibrators 1733-130 and 17458-28204 every 30 minutes for cross
calibration purposes.   The calibration solution was derived from observing the secondary calibrator
1733-130 and then applied to both Sgr A* and 17443-31166.  
This experiment was designed to remove the amplitude
errors resulting from the low elevation of Sgr A* and investigate the
correlated flux variability. 
 The 2006 observation was carried out in the
A configuration using the fast switching technique.  In this
experiment, 7 and 13mm observations were alternated every few minutes using
all the available antennas of the array.  We focus mainly on high frequency
observations because extended free-free emission from ionized gas in the
vicinity of Sgr A* is suppressed at high spatial resolution.

VLBA observations were also carried at 7 and 3mm simultaneous with VLA
observations during 2008 May 5-6, 10-11.  A more detailed description of
these observations will be given elsewhere.  Briefly, these observations (BY122) 
used the inner 8 of the 10 VLBA 
antennas 
excluding the Mauna Kea and St.  Croix antennas which give the longest
baselines that fully resolve Sgr A* due to scatter broadening. 
The observing sequence consisted of
3 minute observations of 1733-130 conducted every hour; 
these served as a
fringe-finder and to calibrate instrumental effects.  Between these scans,
we alternated 5 minute blocks of observations at 7mm (43 GHz) and 3mm (86
GHz).  Each block consisted of a 30 second scan on J1745-2820 followed by a
270 second scan on Sgr A*.  The flux density scale was determined from
standard VLBA antenna gain curves and system temperatures measured during
the observations.

\subsection{Results}

Figure 1 shows four light curves of Sgr A* on 2008 May 5, 10, 6 and 2006
February 10 using the VLA with a sampling time of 3.3sec at 7mm.  
The light curves of the cross calibrated calibrator are shown at the bottom of
each panel in Figure 1. The flux
variation is typically about 0.2--0.6\,Jy with fractional change of 20-30\%
over $\sim$6 hours of observations.  The 2008 May 10 light curve shows  an 
unusual  variation during 8.456h and 8.522h UT. This sharp drop in the flux 
was also simultaneously detected at 13mm  with different set of 
VLA antennas.
The drop in the flux  at radio and submm wavelengths was recently 
discussed  in the context of 
adiabatically cooling plasma blobs that 
are  partially eclipsing the background quiescent emission from Sgr A* (Yusef-Zadeh et al. 2010). 
However,  
we question the reality of this dramatic drop in the flux of Sgr A* by $\sim0.8$ Jy. 
A more detailed account of these measurements 
will be given elsewhere. Excluding the sharp drop in the flux of Sgr A*, a high variation with a 
fractional change of $\sim$40\% is noted on 2010, May 10. The contamination of extended structures in the 
light curve can be significant for baselines less than 100k$\lambda$ so we removed shorter baselines in 
constructing of all the light curves presented in Figure 1a-c (Yusef-Zadeh et al. 2009).  To further minimize any 
contribution of 
extended emission, we have selected data taken with the VLA in its A configuration.  Figure 1d 
shows 
the light curve constructed from the A configuration  observation on 2006 February 10.  
The light curves 
corresponding to 2008 May 10 (excluding the sharp absorption feature) and 2006 February 10 show  flux 
variations of $\sim0.6$ and 0.5\,Jy with mean flux of 1.6 and 2.6\,Jy, respectively.

Figures 2a presents a plot of the structure function of Sgr A* on 2008 May 5 at 7mm. To examine the frequency 
dependence of the variability, the binned CLEAN power spectral density (PSD) 
and the  power spectrum calculated by Uttley et al. (2002) are  presented in Figure 2b,c, respectively. 
To check the CLEAN power spectrum analysis, the power spectrum calculation  using Uttley's technique 
show that  the binned data points in Figure 2b,c  are  consistent with each other but with different 
normalizations. 
These power spectra confirm that most of the power 
is at low frequencies. 
Power-law fits to the structure function   over a 
range between 1 and 10 minutes as well as to the  PSD in Figure 2c 
over the entire frequencies are displayed in Figure 2a,c. 
On short time scales $\tau<$1min the 
structure function and  the PSDs at frequencies between 1 and 10 min$^{-1}$ 
are constant within the error bars. However,  the variability 
at values $\tau>$2min shows a steeper slope consistent with evidence for short time scale variability. 
For comparison, we constructed the structure 
function of the calibrator 17443-31166 and found the slope of the structure function remains flat over all time scales.  
The  flat slopes 
in Figure 2a,b,c  at
 $\tau<$1min 
 are consistent with measurement errors in amplitude or white noise 
(Hughes, Aller \& Aller 1992).
For flux variability that is simply due to measurement errors with standard error $\sigma$, it is expected that 
structure function amplitude is 2$\sigma^2$.

%Most of the power in the time variability of Sgr A* appears to be on time scales $\tau>$60min with a turnover at 
%time scales exceeding $\sim$100min.   
%This turnover is considered to be characteristic time scale beyond which the 
%variability is uncorrelated.  The characteristic time scale is interpreted to be due to multiple flaring events in 
%central engines of AGNs 
%Kataoka et al. (2001) (see Emmanoulopoulos et al. 2010). 

To examine the reality of the short time variability at
$\tau>$1min, 
we have also constructed 
autocorrelation functions of the light curves taken on 2008, May 5. 
Figure 2d  shows a  plot of the autocorrelation function of the measurements 
showing  a peak at zero time lag,  followed by a 
sharp  drop implying a lack of correlation on short time scales. 
The steep drop in the autocorrelation function is consistent with short time scale variability 
inferred from SF and PSDs.

%The increase in the amplitude of the autocorrelation 
%on hourly time scales is consistent with correlations in the flux  variability,  as seen 
%in the slow variability noted in  Figures 1a and 2a,b.c. 

To  check the validity of the analysis of  the data taken on 2008 May 5, 
we carried out simultaneous VLBA and VLA  measurements 
at 7mm. 
Due to the large interstellar scatter  broadening  of Sgr A*, only few short baselines 
of VLBA  antennas 
could  be used to investigate the nature of the variability of Sgr A* at 7\,mm.  The same plasma is responsible for 
daily  time variability at centimeter wavelengths
(Macquart and Bower 2006). 
Scintillation models predict variability at a level of 
10\% at 7mm wavelengths on a time scale of few days 
(Macquart and Bower 2006). Thus, the variability measurements presented 
here show large flux variability on much shorter time scales, thus, 
should not be affected by interstellar scintillation. 
Figure 3a-d show the SF, two different  PSDs and the autocorrelation function of VLBA data at 7mm, 
all of which are consistent with short time scale variability, as measured by the VLA and shown 
in Figure 2.  
The measurements error bars in Figure 3  
are high and the flat slope fitting the short time scale variation is 
consistent with high measurements errors.  
The consistent results from VLA and VLBA measurements on short time scales 
support the suggestion by earlier studies that 
the radio variability of Sgr A* arises from the inner AU of Sgr A* 
(Yusef-Zadeh et al. 2009) and that both show similar time scale for 
minute time variability. 

%The turnover time scale $\sim$100\,min is similar to that of VLA 
%measurements.  

Figures 4a-d show the SF, PSDs and the autocorrelation function of VLA  data 
taken on 2008 May 10 at 7mm.  The 
structure function  rises at short time scales,  plateaus between 5 and 100\,min, before rising again 
at  longer time lags.  The rapid steepening of the structure function above 300\,min reflects the decline 
of  the light curve over the duration of the observations, and the sharp drop apparent at on the very longest 
time scales is an artifact of the very limited sampling available at the longest lag times.  It is clear 
that the slope of the structure function changes between minute and hourly time scales.  Similar behavior 
has been observed in the structure function plots of X-ray data taken toward AGNs 
(Kataoka et al. 2001).   The  slopes to  the CLEAN PSD and Uttley's PSD, as shown in Figure 4b,c,  
range between $\alpha=-0.65$  and --0.86 and  are consistent within 3$\sigma$ errors of each other. 
Power law fits to  the SF at short  time scales between 0.1 and 1\,min and 
 to  the PSDs over their  entire frequency spectra as well as the  sharp drop in  
the autocorrelation function at 0 time lag  
support evidence of short time scale variability. 

%The best-fitting power law slopes  are displayed and fall somewhere between white noise and flicker noise. 

The structure function for the 2008 May 10 light curves, as shown in Figure 4a, differs in two ways from that for 
2008 May 5.  First, a number of dips in the structure function show what at face value might be considered as QPO 
activity. A dip is produced 
when a pair of measurements separated by a time lag $\tau$ have similar flux density.  The strongest dip is seen 
around 20 minutes followed by weaker dips at 50 and 100 minutes.  Similar dips are also noted in the structure 
function plot at 13mm which were taken simultaneously with 7\,mm.  These dips which are not detected in other three 
days of observations in May 2008 are caused mainly by the sharp drop in the flux of Sgr A*, as shown in Figure 1c. If 
we remove the sudden drop of flux, the dips in the SF  disappear. Thus, we can not confirm the reality of these 
dips.

The second feature apparent in Figure 4a is the steepening of the slope of the structure function at $\tau<$3 min 
down 
to time scales as short as 0.1\,min, indicating that there is intrinsic variability on very short time scales.  The 
reality of this short time scale variability is strengthened by the flat structure function of the calibrator.  In 
addition, we note that the expected value of the structure function once it becomes dominated by measurement error is 
$1.5\times10^{-3}$\,Jy$^2$.  The amplitude of the structure function with lag time greater than 0.1 min is higher 
than 
this, supporting the reality of such short time scale variability.  These measurements suggest continuous variability 
from sub-minute time scale to  hour time scale variability, but with varying slopes. Figure 4d shows a 
dramatic drop in the amplitude of the autocorrelation during the initial ten minutes which is 
consistent with the PSD not flatteining at higher frequencies.  The steep drop in 
the autocorrelation 
function  supports the intrinsic minute time scale variability.

Figures 5a-c and 6a-c show the SF, CLEAN PSD and the autocorrelation function of VLA data taken on 2008 May 6 at 7 and 13mm, 
respectively. Figures 5a and 6a show the plots of the structure function for the 2008 May 6 light curves at 7 and 13mm, 
respectively.  Overall, the structure function is a shallow power-law at short time scales before steepening at long time 
scales.  Although the measurement errors of the 13mm data are greater than at 7mm, the structure function at 13mm has a 
similar trend to that seen at 7mm.  The power-law indices at 7 and 13mm at short time scales are 
0.22$\pm3\times10^{-4}$ 
and 0.32$\pm2.7\times10^{-4}$ 
 and and at 
long time scales are 1.39$\pm5.9\times10^{-5}$ 
 and 1.20$\pm3.1\times10^{-4}$, respectively.  The transitions from a shallow to a steep slope of the structure 
function are at $\sim30$ and 60 
minutes at 7 and 13mm, respectively. The constructed CLEAN PSD and the autocorrelation of the 2008 May 6 data at 
7 and 13mm show the slopes of the power spectra are not flat and that there is a steep drop in the amplitude of the 
autocorrelation function. These figures all show  the evidence for short minute time scale variability at radio 
wavelengths.

Figures 7a-c  and 8a-c show the SF, CLEAN PSD and the autocorrelation function of VLA  data 
taken on 2006 February 10 at 7 and 13mm, respectively.
 As pointed out earlier, these measurements 
should have no contamination from the  extended emission 
from the ionized gas surrounding Sgr A*. 
These figures show collectively the same pattern of continuous
variability from 0.1 to 250 min.  
 The sub-minute time scale variability and the
deviation from a flat slope at short time scales in  the SF and high frequencies 
are consistent with the short time scale variability of Sgr A*. 
The smoothly varying slope in these high resolution observations 
is consistent with other low resolution measurements taken in the  C configuration of the VLA. 
Furthermore, the steep falloff of the correlation at short time scales
implies  rapid and continuous fluctuations over a wide range
of time scales in the flux of Sgr A*.  

%The best constraint, 
%we can place on
%the turnover time scale is $\tau\sim200-300$ \,min at radio wavelengths.

\section{IR and X-ray Variability}

Structure functions have also been calculated for several nights of Keck
observations of Sgr~A* (Do et al. 2009).  We have used data taken with VLT,
 XMM-Newton and VLA
observations on 2007, April 4 at IR, X-ray and radio wavelengths (Porquet et al. 2008; Dodds-Eden et al. 2009; 
Yusef-Zadeh et al. 2009).
Figure 9a-f show the structure function and the corresponding CLEAN  PSD plots of IR, X-ray, radio
observations, respectively, sensitive to time lags ranging from about 0.5 to a few
hundred minutes. 
The X-ray and IR data on 2007 April 4 revealed 
simultaneous bright X-ray and IR flares at the beginning of observations. 

The power-law fit to the IR data implies slope of 0.9 in the
structure function.  Different nights of Keck
observations (Do et al. 2009)  show value of $\beta$ varying between 0.26
and 1.37 with lag times ranging between 1 to 40 minutes.  The value of
$\beta$ from VLT measurements are consistent with Keck measurements. 
We also note evidence of 
variability at about  minute time scale at IR wavelengths (see also the analysis
by Do et al. (2009)  and Dodds-Eden et al. (2009).  The slope of the CLEAN PSD at high frequencies 
is consistent  with the short time scale variability inferred from SF analysis. 
There is no evidence for
QPOs in the time domain that was searched.  The reality of QPO activity of
a hot spot orbiting Sgr~A* is hotly debated mainly because of the low
signal-to-noise and possible intermittent nature of such behavior 
(Eckart et al. 2006; Meyer et al. 2008; Do et al. 2009).

Unlike the IR structure function which is fit  by a single power law,
X-ray structure function and CLEAN PSD  give different
characterization of the variability of Sgr~A* in X-rays.  We note a rise
 at short time scales, similar to that of IR, though somewhat shallower,
followed by a plateau
with time lags ranging between 30 and 300 minutes before a steepening of
the structure function again at longer time scales. 
 Due to limited
sensitivity and 100-second sampling of X-ray data, the flat part of the SF
with time lags of few minutes indicate that there is no minute time scale
variability and that the emission is dominated by white noise.  The plateau
time lags range between 30 and 300 minutes and is also consistent with 
white noise where there is no correlation of signals.  
At time lags greater
than 300 minutes, the correlation begins again. 
The steepening of the structure
function at 300 min is due to the large count rate difference between
the beginning of the light curve when there was a bright X-ray
flare and the end of the observation when the emission was at its quiescent
level.  The X-ray shape of the structure function plot of Sgr A* is
remarkably similar to that of Mrk 421 with time lags that are an order of
magnitude larger (see Figure 4c of Kataoka e al. 2001).  
Similarly, the
mass of the black hole in Mrk 421 is estimated to be 50 times higher than
the mass of Sgr A* (Barth, Ho and Sargent 2003).  The characteristic time
scale of X-rays from Mrk 421 is considered to place a constraint on the
size of the variable X-ray emission from the base of the jet 
(Kataoka et al. 2001). 

Figures 9e,f  represents the structure function and PSD at  7mm  taken simultaneously with IR
and X-ray data, as part of an observing campaign that took place on 2007, April 4
(Yusef-Zadeh et al. 2009). 
 We note a  rise of the amplitude of the SF, 
 similar to that seen in the IR structure 
function plot. The power-law fit
to the radio data shows slope of 0.78 in the
structure function which is close to  that of IR data, as seen in Figure 9a.
A correlation between the optically thin IR, 
X-ray flare and optically thick 7mm radio flare has been suggested for the strong flare 
that occurred on this day (Yusef-Zadeh et al. 2009; Dodds-Eden et al. 2009). The 
radio flare
emission at 7mm was argued to be delayed with respect to the near-IR and X-ray flare
emission, consistent with the plasmon picture. The similar behavior of the structure
function of IR and radio data is not inconsistent with a picture that flaring activity in
radio and IR wavelengths is correlated.

\section{Discussion}

\subsection{Long Time Scale Variability}

Our multi-wavelength monitoring of Sgr A* characterizes the intrinsic time variability of Sgr~A* by studying the 
structure function of the observed light curves.  Structure function analysis of the IR, X-ray and radio data 
suggests  
that most of the power falls in the long time scale fluctuation of the emission from Sgr~A* and that the variation is 
generally aperiodic with no obvious QPO activity, confirming earlier analysis of IR data 
(Do et al. 2009; Meyer et al. 2008). 
  The structure function analysis of radio data 
shows a number of new features, the most interesting of which is a statistically significant time 
variability on subminute to hourly time scales.  Unlike the IR structure function, which can be 
well-represented by a single power law, at radio wavelengths the structure functions are more complex and 
could only be fit by multiple power law components.

Using the long time scale lags, we fit a power law to the PSD and SF of radio data and find the power indices of 
radio variability. 
The long time scale variability of Sgr A* at 
radio wavelengths characterized in the SF plots is  similar to the duration of typical flares as detected in both 
radio and submm wavelengths  
(e.g., Yusef-Zadeh et al. 2006; Marrone et al. 2006).
 These measurements at 7 and 13mm are consistent 
with the power spectrum analysis of the time variability of Sgr A* suggesting intraday variability at 3mm 
(Mauerhan et al. 2005). 
The rapid decay of the SF plots at large lag times appears to  reflect a turnover in the PSD of the 
variability.  However, this turnover is probably due to an artifact of   limited sampling of radio data at large time 
lags  (see Emmanoulopoulos et al. 2010). 

The $\sim$ hour-long flaring in radio and sub-mm has been argued to be due
to adiabatic cooling of synchrotron-emitting electrons in an expanding
plasma blob (Yusef-Zadeh et al. 2009).  These measurements  support 
this picture and show that the light curves peak at successively lower
frequencies (submm, millimeter and then radio) as a self-absorbed
synchrotron source region expands after the initial event that energizes
the electrons.  The emission at a particular frequency peaks as the blob
becomes optically thin at that frequency, so the blob size determines the peak
flux of the flare and the expansion speed determines the flare duration.
The estimated expansion speed of the plasma is a few percent of $c$,
the plasma itself may be bound to Sgr A* or be embedded in the base of
a jet (Maitra et al. 2009; Yusef-Zadeh et al. 2009). 
In this scenario, the contributions of the flares to the structure
functions at 7\, and 13\,mm should be related to one another.  To examine
this, first consider the frequency-dependence of the properties of
individual flares.  For an $E^{-p}$ electron energy spectrum the light
curve of each flare follows the characteristic frequency-dependence of the
plasmon model (van der Laan 1966; Yusef-Zadeh et al. 2006):
\begin{equation}
    S_\nu = S_0 \, \left(\frac{R}{R_0}\right)^3 \, \left(\frac{\nu}{\nu_0}\right)^{5/2} \, \frac{1-e^{-\tau_\nu}}{1-e^{-\tau_0}} \,,
    %\label{eq:Snu}
\end{equation}
where the optical depth of the blob
\begin{equation}
\tau_\nu = \tau_0 \, \left(\frac{R}{R_0}\right)^{-(2p+3)} \, \left(\frac{\nu}{\nu_0}\right)^{-(p+4)/2} \,,
    \label{eq:taunu}
\end{equation}
and the reference optical depth $\tau_0$ is chosen so that the peak flux at
frequency $\nu_0$ is $S_0$ and this occurs when the blob radius is $R_0$.
This means that $\tau_0$ is determined by $d\,S_\nu/dR=0$, it turns out
that $\tau_0\sim 1$ and is a weak function of $p$ (Yusef-Zadeh et al. 2006).  At
frequency $\nu$ the peak flux
\begin{equation}
    S_p = S_0 \left(\frac{\nu}{\nu_0}\right)^{(7p+3)/(4p+6)}
    \label{eq:Sp}
\end{equation}
occurs when the blob radius is
\begin{equation}
    R_p = R_0 \, \left(\frac{\nu}{\nu_0}\right)^{-(p+4)/(4p+6)} \,.
    \label{eq:Rp}
\end{equation}
Then we may rewrite the flux and optical depth at $\nu$ as
\begin{equation}
    S_\nu = S_0 \, \left(\frac{R}{R_p}\right)^{(7p+3)/(4p+6)} \, \frac{1-e^{-\tau_\nu}}{1-e^{-\tau_0}}
    \label{eq:Snu2}
\end{equation}
and
\begin{equation}
\tau_\nu = \tau_0 \, \left(\frac{R}{R_p}\right)^{-(2p+3)} \,,
    \label{eq:taunu2}
\end{equation}
respectively.  These expressions show that the flux as a function of blob
radius at $\nu$ can be found from the light curve at $\nu_0$ through a
simple linear stretch of the $R$-axis by a factor of $R_p/R_0$ and a
compression of the flux axis by a factor $S_p/S_0$.  Under the assumption
that the blob has constant expansion speed, the mapping between $R$ and
time is linear and the light curves behave in the same way.  To summarize,
the adiabatic expansion scenario implies that the amplitude and time scales
of a single flare scale as $\nu^a$ and $\nu^{-b}$ respectively, where
\begin{equation}
    a = \frac{7p+3}{4p+6}
    \label{ea:a}
\end{equation}
and
\begin{equation}
    b=\frac{p+4}{4p+6}\,.
    \label{eq:b}
\end{equation}

Suppose now that the variations in the light curves of Sgr A* at 7\,mm arise
through a superposition of flares, each with the same $E^{-p}$ electron
energy spectrum but with a distribution of amplitudes and time scales.
Because the amplitude and time scales of the contribution of each flare
scale with frequency as $\nu^a$ and $\nu^{-b}$ respectively, the structure functions of
the light curve at $\nu$ and $\nu_0$ are related by
\begin{equation}
    SF_\nu(t) = \left(\frac{\nu}{\nu_0}\right)^{2a} \, SF_{\nu_0}\left(\,(\nu/\nu_0)^b\,t\,\right)\,,
    \label{eq:SFnu}
\end{equation}
where $a$ and $b$ are given by equations 7 and 8.
In particular, note that if the
structure function at $\nu_0$ is a power-law, the
structure function at $\nu$ will also be a power law with the same index but
different normalization.

To test this hypothesis, we take the power-law fits to the structure
functions at 7\,mm for 2008 May 6 and 2006 February 10 and compute the
13\,mm structure function predicted by eq (\ref{eq:SFnu}), setting $\nu_0$
and $\nu$ to 43 and 22\,GHz respectively.  
The amplitude of the predicted 13mm SF using our model 
and extrapolated from the 2008 May 06 data at 7mm is lower by 30\%  than that observed at 
13mm. 
%times $\tau < 30$min is lower by 30\% than the observations. 
The only adjustable parameter
available is $p$, the power law index of the electron energy spectrum.  We
overplot the result on the measured 13\,mm structure functions in Figure 10.
we conclude that structure functions at 13\,mm for 2008 May 6 and 2006
February 10 are quantitatively reproduced from their 7\,mm counterparts for
$p \sim 0.5-1$. This is  consistent  with earlier fits to individual large flares in
other radio and sub-mm data sets (Yusef-Zadeh et al. 2006, 2009).

\subsection{Short Time Scale Variability}

The most interesting feature of the analysis presented here is that the
structure function at radio wavelengths shows short time scale variability
of Sgr A* $\sim$0.3\,min with a shallower slope than seen at longer time
lags.  This is the shortest time scale variability that has been detected
toward Sgr A*  and suggests that radio emission may arise from the
innermost region of the accretion flow.  The best case for minute time
scale variability is seen in Figure 3a and 5 where the structure function
rises for lag times greater than 0.3 minutes.  In both cases, the amplitude
of the structure function at time scales greater than 0.3\,min for data
taken on 2008 May 10 and 2006 February 10 is greater than twice the mean
square of the measurement errors, providing  the evidence  for subminute
time scale variability.  
The transition time scales where the slope of the SF changes was estimated
from 30 and 60 minutes at 7 and 13mm.  These transitions suggest that the
physical mechanism for production of the variability may be different.  The
different lag time at 7 and 13mm can be accounted for in terms of optical
depth effect of hourly time scale variability in an adiabatic expanding
synchrotron source.  Previous time delay determinations based on individual
identified flare events show a time delay of 20-40 minutes between the
peaks of emission at 7 and 13mm
(e.g., Yusef-Zadeh et al. 2009)
as well as
the increase of  the duration of the flare emission in the time domain at
lower frequencies.  The latter effect is likely to be responsible for the
transition in the slope seen in the structure function analysis.

%In addition, the short time scale variability is
%evidenced best when Sgr A* shows a high mean flux.

As noted before, the component of the variability shows a different slope
at short time scale than that at hourly time scale which dominates the
power spectrum of the variable emission.  The steepening of the slope at
longer time lags reflects the systematic increase of the flux from Sgr A*
over the course of the observations, and is removed if a linear increase in
flux is subtracted from the light curve.  After subtracting off the secular
change over several hours of observations, the structure function shows
that there is continuous variability at 7mm on minute to hourly time scale
and that there is variability on a longer time scale than several hours.
Given that hourly time scale variability is interpreted in the context of
adiabatically expanding hot plasma blob with sub-relativistic expansion, it
is natural to consider that the short time scale variability is also
optically thick.

Sub-minute time scale variability places a strong constraint on the size of
the emission region.  Light crossing time arguments place an upper limit
comparable to the Schwarzschild radius of Sgr~A*, i.e.\ $\approx 0.1$ AU
for a Galactic center distance of 8\,kpc (Doeleman et al. 2008).
The flux variation on minute time scales at 7\,mm is $\sim$10-30 mJy.
Assuming that this is produced by adiabatic expansion of a uniform
synchrotron source with an $E^{-1}$ electron spectrum, the optical depth at
the peak $\tau_0 \approx 0.95$.  Assuming that the electron spectrum
extends between 1 and 100\,MeV and that there is equipartition between the
magnetic field and the electrons, then a peak flux of 30\,mJy is obtained
for $R_0 \approx 1.1\times 10^{12}$\,cm and magnetic field strength
$B_0\approx14$\,G. The flare time scale $R_0/v$ is 60 seconds for an
expansion speed $v \approx 0.62\,c$.  For a peak flux of 10\,mJy we obtain
$R_0\approx0.67\times10^{12}$\,cm, $B_0\approx16$\,G and $v\approx0.37\,c$
respectively.  Clearly this model would have difficulty producing 30\,mJy
flux variations on 0.3\,min time scales, but 10\,mJy fluctuations would be
viable.

Associating a proton with each synchrotron-emitting electron, the mass of
the source region for the 30\,mJy case is $\sim 3\times 10^{18}$\,g.
Continual variations on a 60 second time scale then imply a processing rate
of material $6\times 10^{-10}$ \msol yr$^{-1}$.  The energy in
magnetic fields and relativistic electrons in the expanding source region
declines on the expansion timescale, implying a transfer of energy into the
immediate surroundings at a  rate $(\frac{4}{3} \pi R_0^3) (2B^2/8/pi) /
(R_0/v) \approx 400L_\odot$.  These rates will be somewhat higher if
their are multiple overlapping flares at any given time.

The origin of these short time fluctuations is unclear, but the inferred
explosive expansion of the emitting plasma at near-light speeds is
suggestive of that they might feed into an outflow or jet.  We note that
the inferred scale length corresponding to one-minute light travel time is
comparable to the time averaged spatially resolved 0.1AU scale observed
at 1.3mm by (Doeleman et al. 2008).  The quiescent variable emission from
Sgr~A* could then be interpreted mainly as an ensemble average of numerous
flares that are detected on minute-time scale.  This short time scale
emission or quiescent variability could be due to fluctuations in the
accretion flow of Sgr~A* due to magnetic field fluctuations resulting from
MRI, as recent MHD simulations in a number of studies indicate.

\section{Conclusions}

In conclusion, the rapid fluctuation of emission from Sgr~A* allows us to
probe the spectacular activities of the central engine at remarkably small
spatial scales.  
 We have
constructed structure functions, power spectra and autocorrelation functions
using radio, IR and X-ray data in order to
characterize the time variability of Sgr A*.  These plots indicate that
most of the power in the time variability is on hourly time scales.
However, the shapes of the structure function in X-rays and radio
wavelengths are different than that of IR data. 
Radio continuum
variability is detected to be continuous from short subminute time scale to
long hourly time scale.  We argue that rapid fluctuations of the radio
emission imply rapid expansion that could feed the base of an outflow or
jet in Sgr A*. The bulk of the continuum flux from Sgr A* at radio and submm  wavelengths 
is believed to
be generated in its  accretion disk. The localization and characterization 
of the source of
variable continuum emission from Sgr A*  give us opportunities to further  our understanding 
of the launching and transport of energy in the nuclei of Galaxies.

% implying that X-ray and
%radio variable emission is reprocessed IR emission.  

\acknowledgments
This work is partially supported by grants AST-0807400
from the National Science Foundation and DP0986386 from the Australian
Research Council.  We thank Mark Reid for his help in reducing VLBA data.

\vfill\eject 

\begin{figure}
\center
\includegraphics[scale=0.35,angle=0]{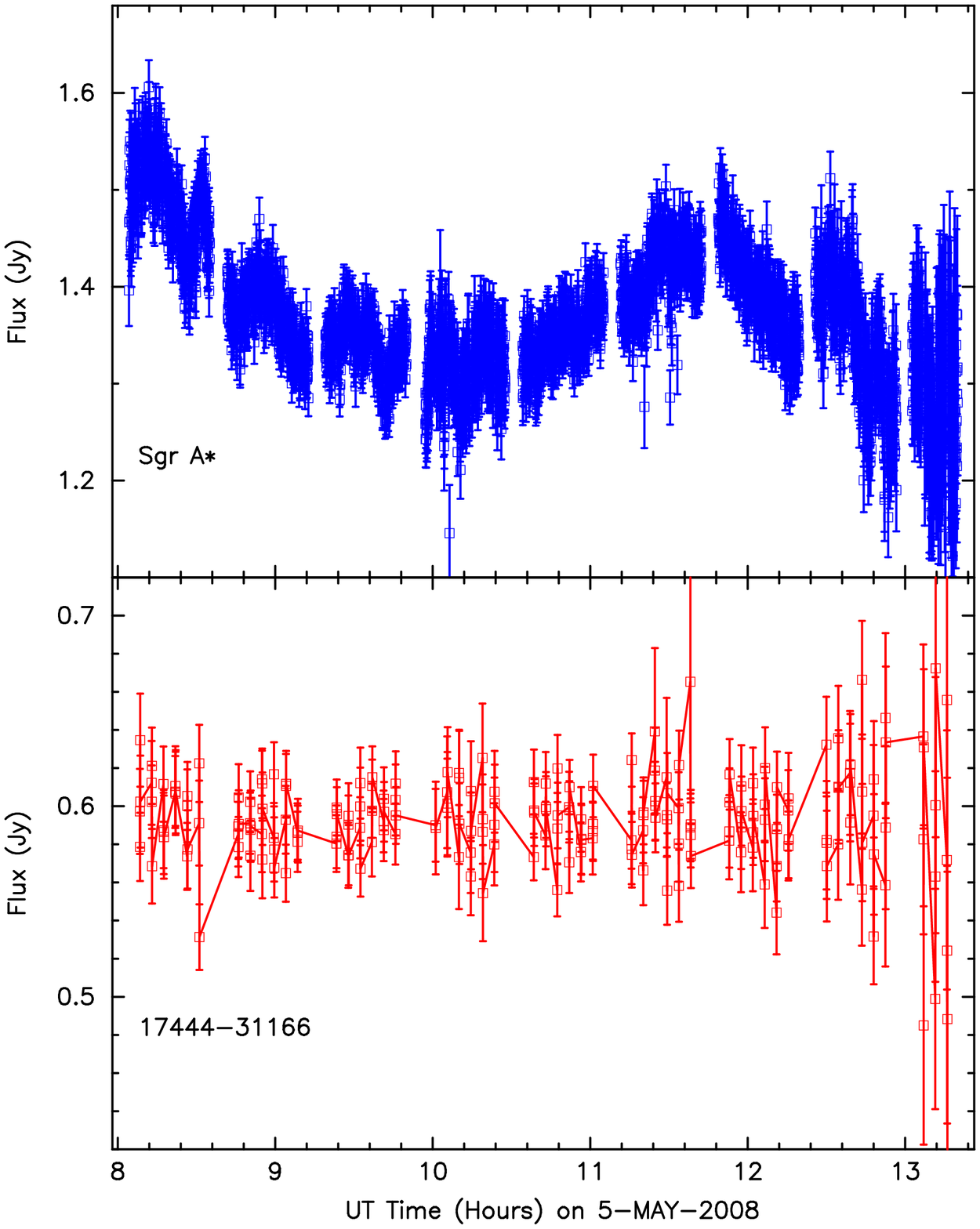}
\includegraphics[scale=0.35,angle=0]{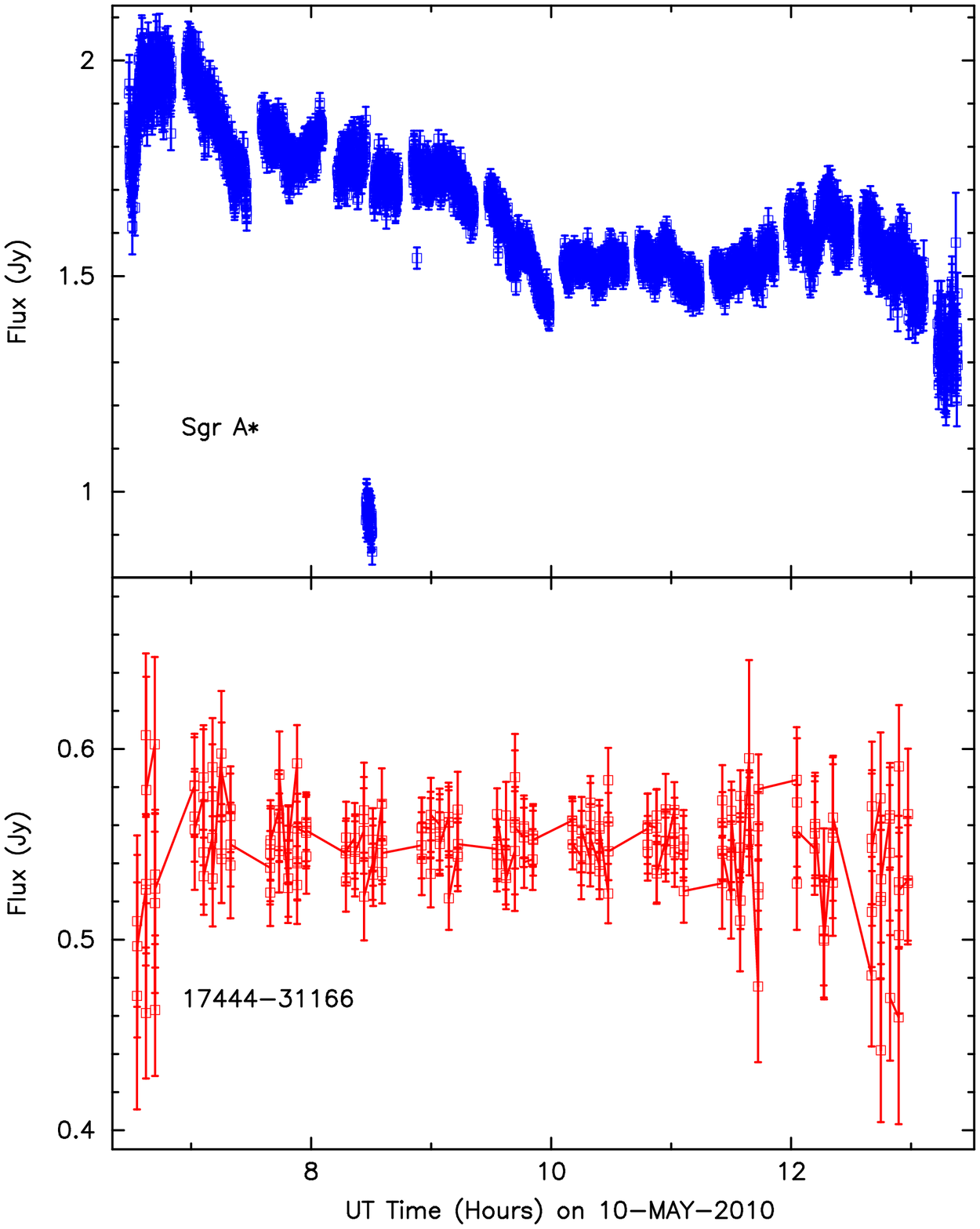}
\includegraphics[scale=0.35,angle=0]{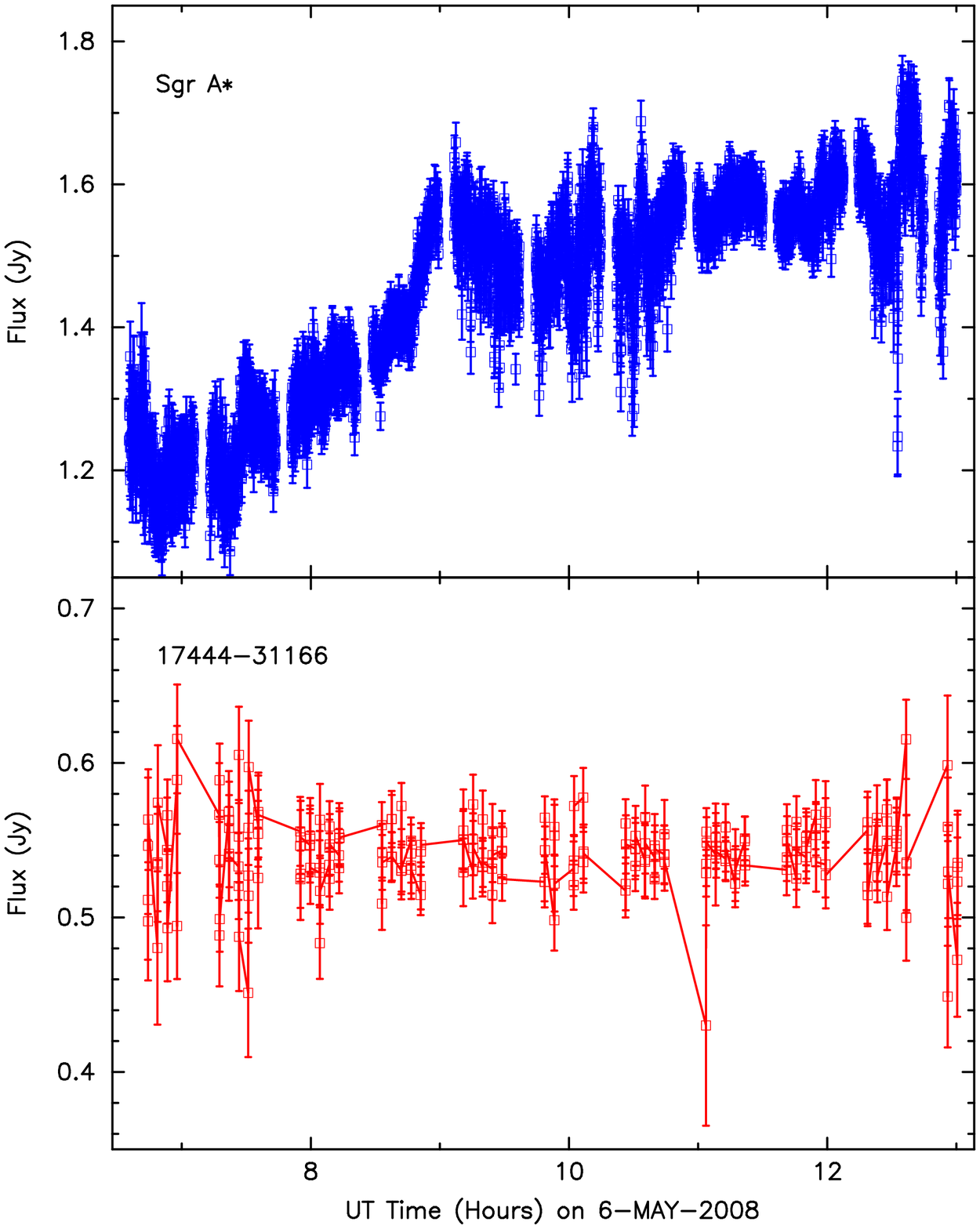}
\includegraphics[scale=0.35,angle=0]{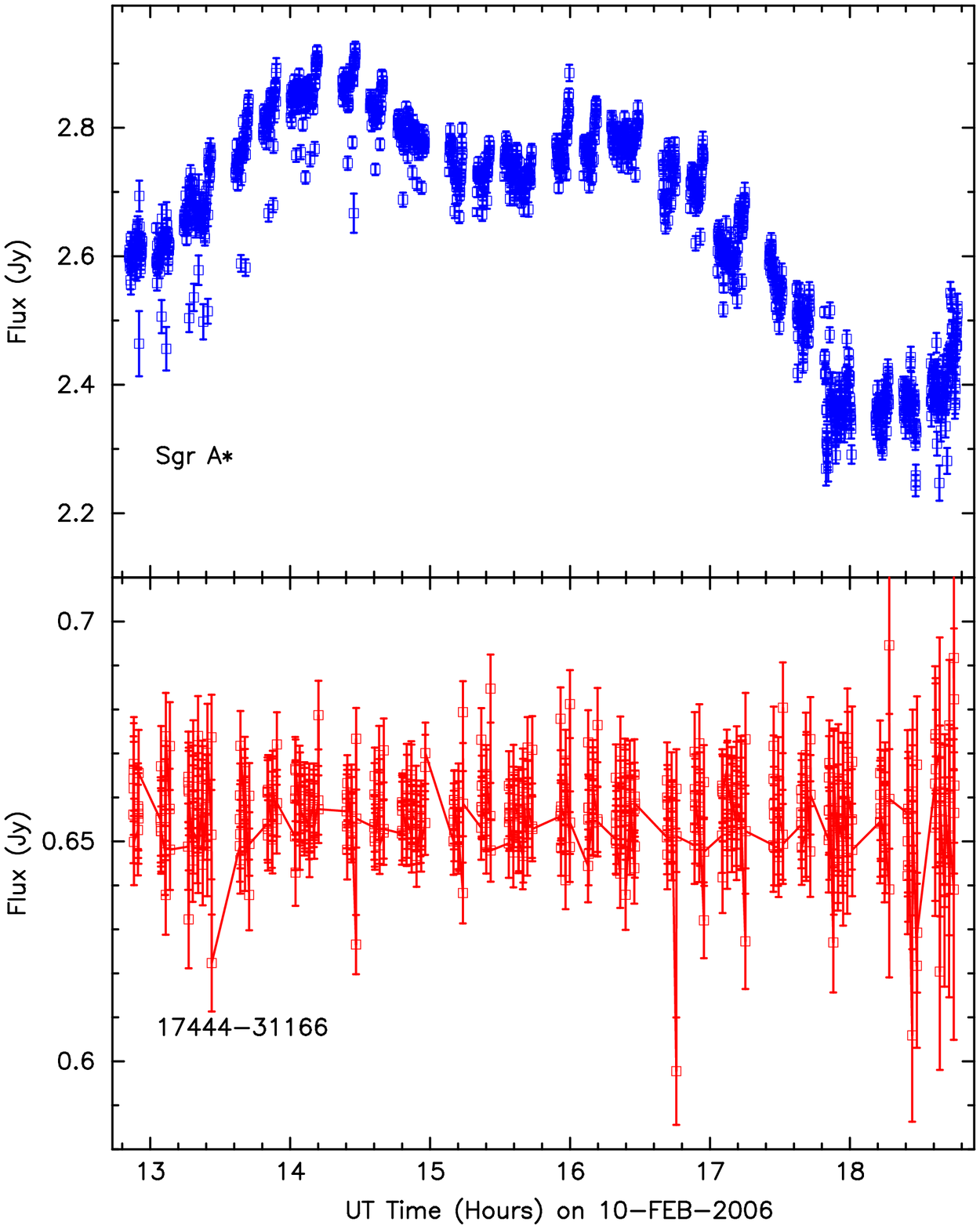}
\caption{
{\it (a) Top Left}
A  light curve of Sgr A* at 7mm for
2008 May 5 observations with a 3.3 sec sampling time and
the corresponding calibrator 17444-31166.
{\it (b) Top Right}
Same as (a) except on 2008, May 10.
{\it (c) Bottom Left}
Same as (a) except on 2008, May 6.
{\it (d) Bottom Right}
Same as (a) except on 2006, February 10.
}
\end{figure}

\begin{figure}
\center
\includegraphics[scale=0.35,angle=0]{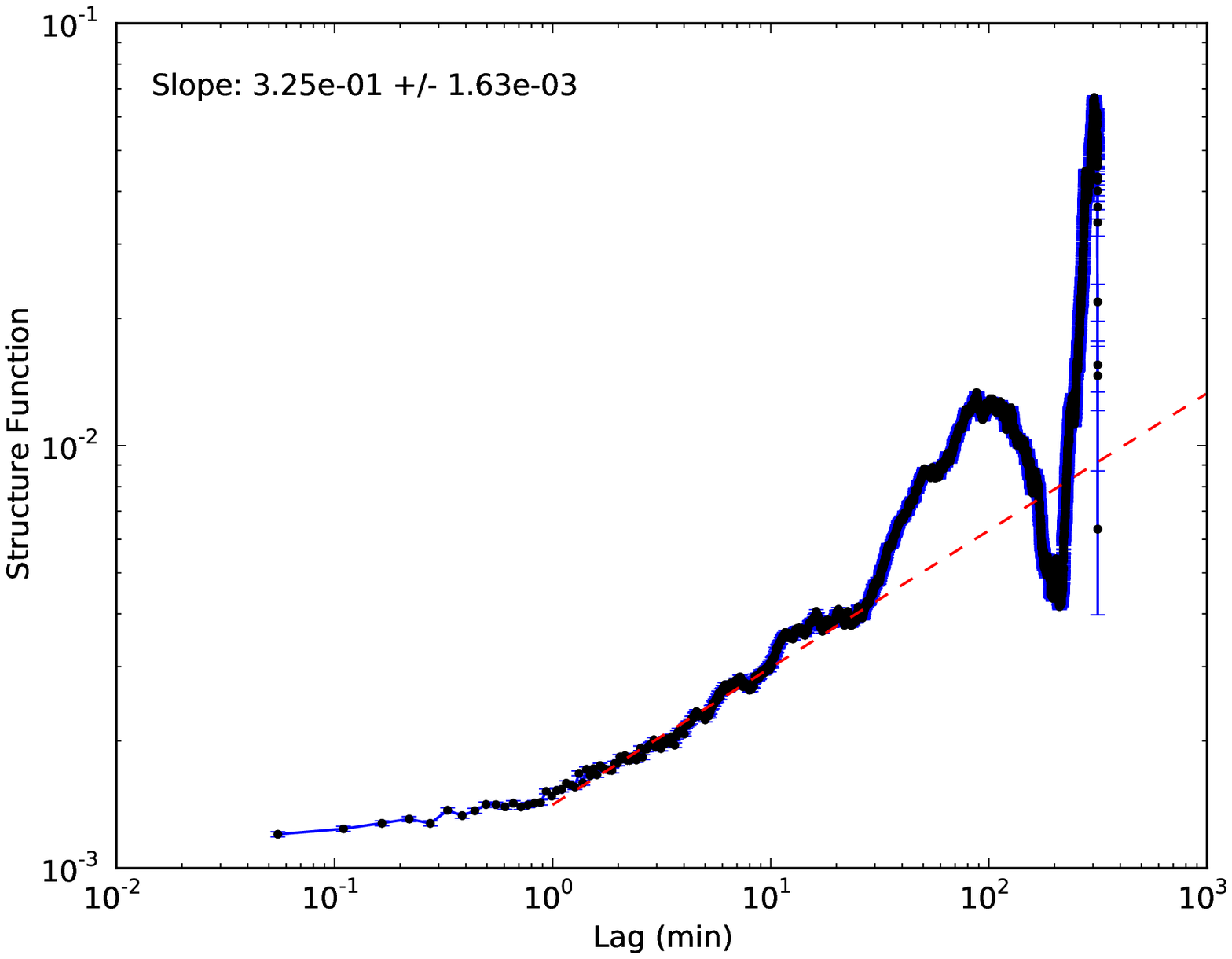}
\includegraphics[scale=0.35,angle=0]{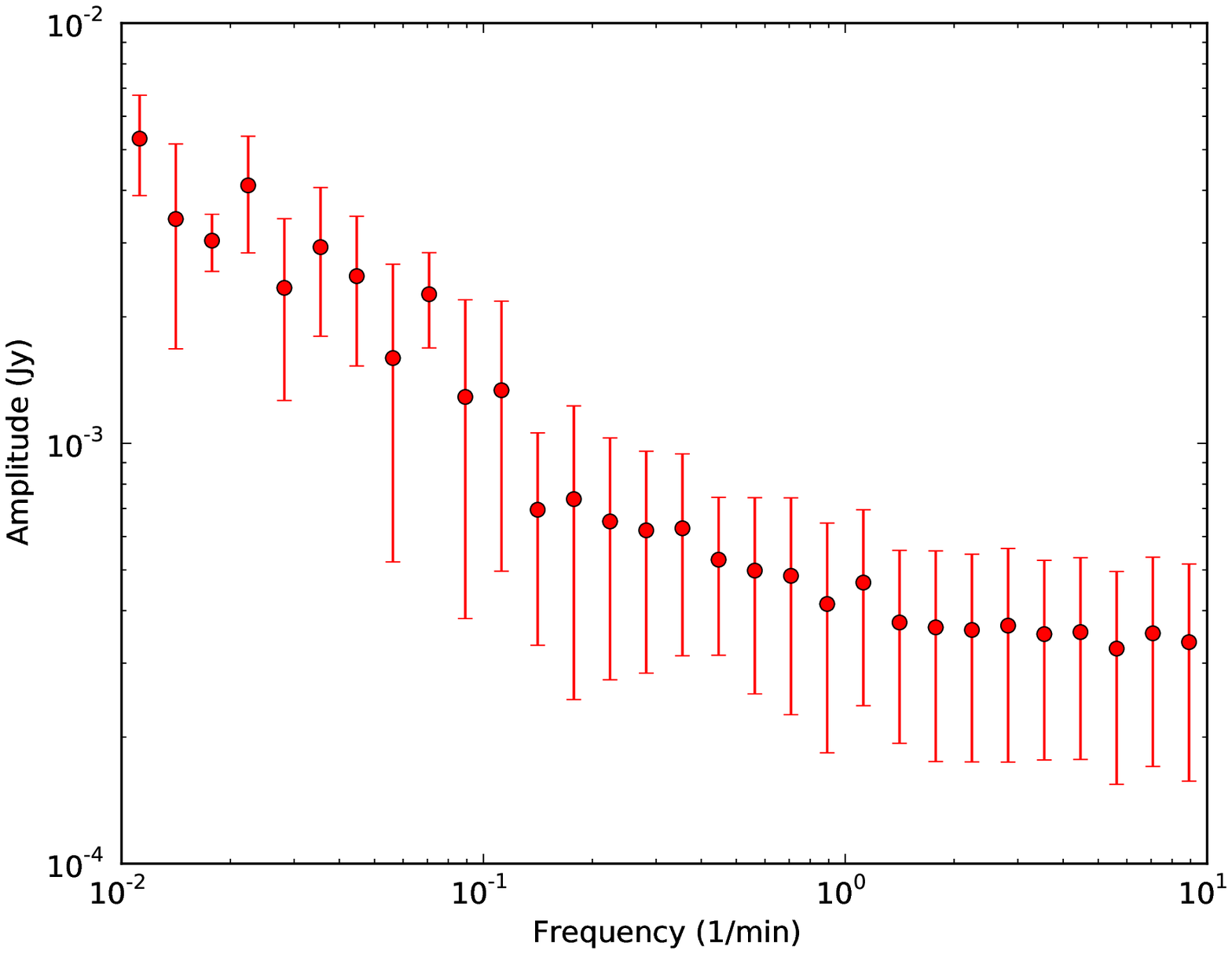}
\includegraphics[scale=0.35,angle=0]{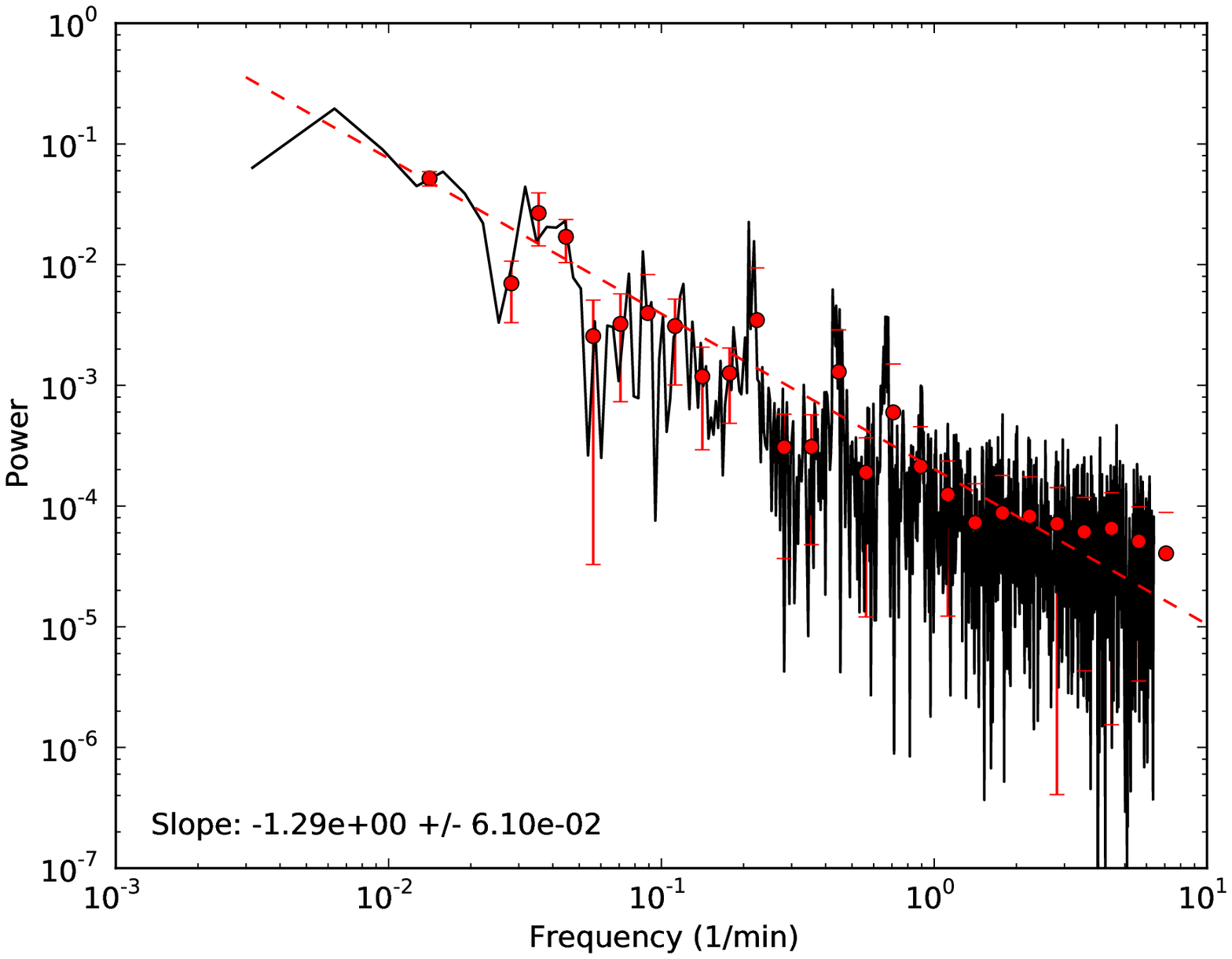}
\includegraphics[scale=0.35,angle=0]{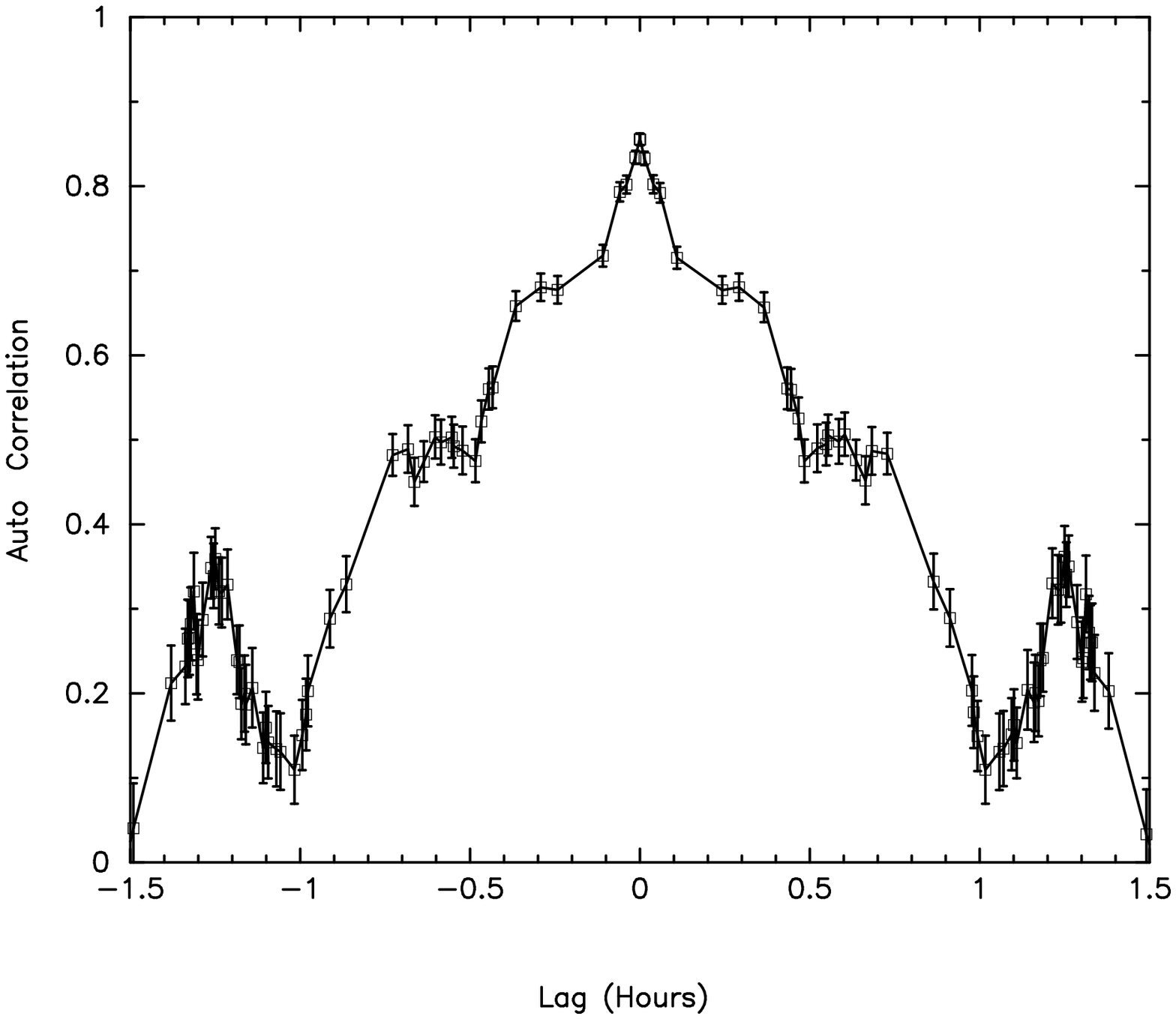}
\caption{
{\it (a) Top Left}
A structure function plot (units Jy$^2$) for VLA observation of
Sgr~A* at 7mm on 2008, May 5 with data sampling of 3.3 sec.
The mean error of the light curve is $\sigma$=0.026 Jy.
The squares of the mean measurement
errors are 6.6$\times10^{-4}$ Jy$^2$ at 7mm.
{\it (b) Top Right}
The CLEAN PSD of data shown in (a). The red dots represent  smoothed bins. 
{\it (c) Bottom Left} The PSD using Uttley's technique with red dots show the binned data. 
The slope fitted through the entire spectrum is displayed. 
{\it (d) Bottom Right} 
A plot of the autocorrelation function.
Power-law fits to the structure function   over a 
range between 1 and 10 minutes as well as to the corresponding  PSD, shown in  
(c),  over the entire frequencies are displayed on each  figure.  
}
\end{figure}

\begin{figure}
\center
\includegraphics[scale=0.35,angle=0]{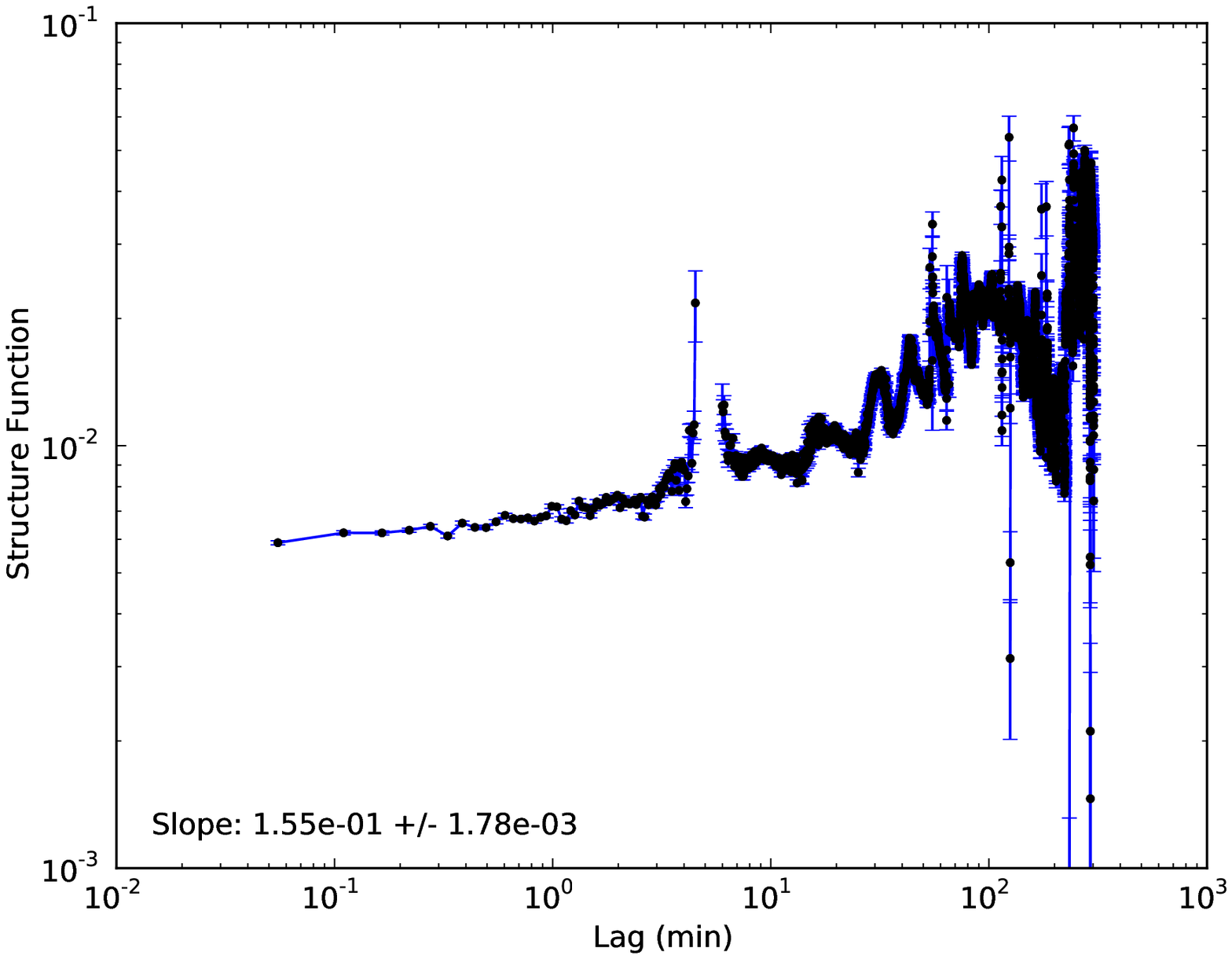}
\includegraphics[scale=0.35,angle=0]{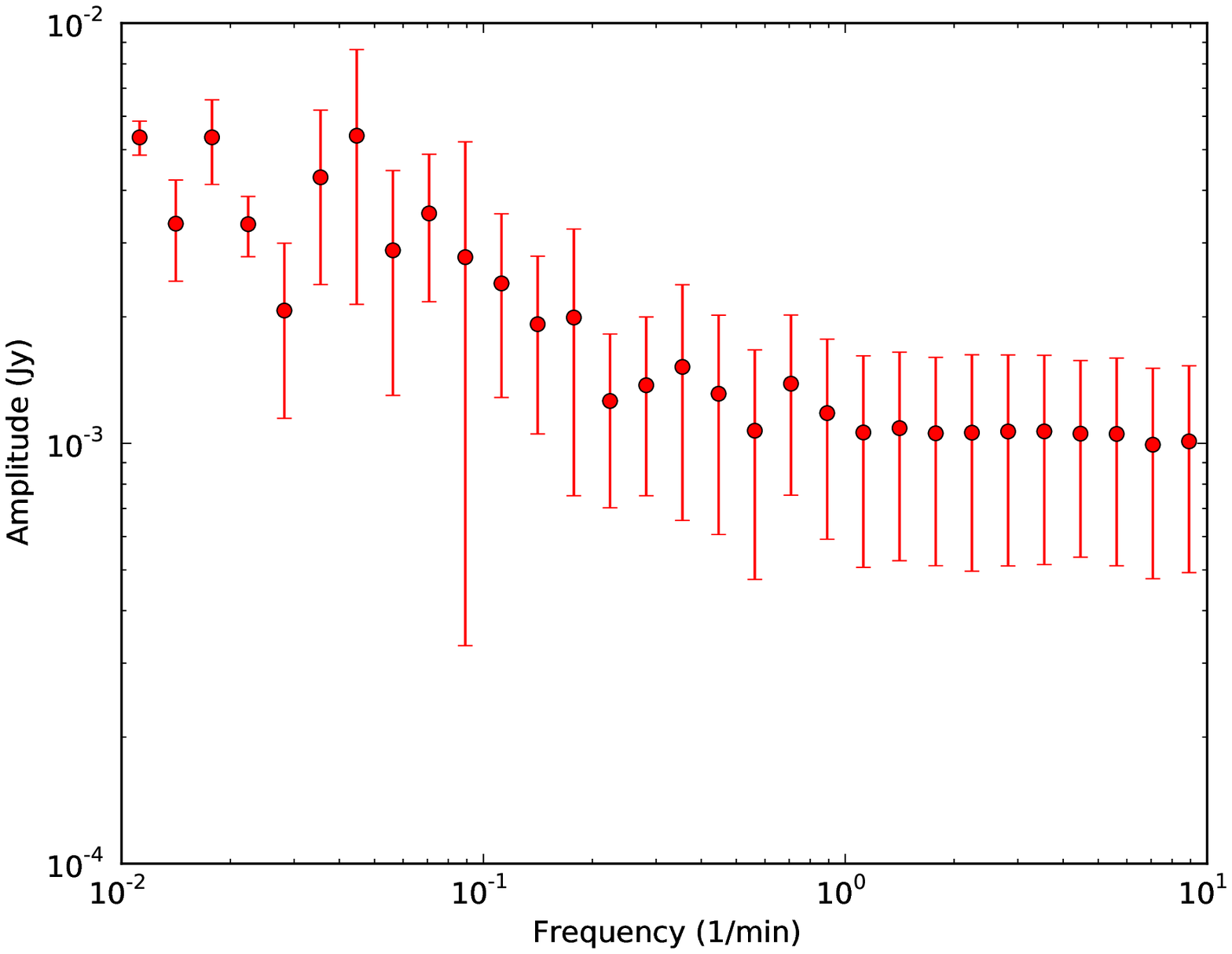}
\includegraphics[scale=0.35,angle=0]{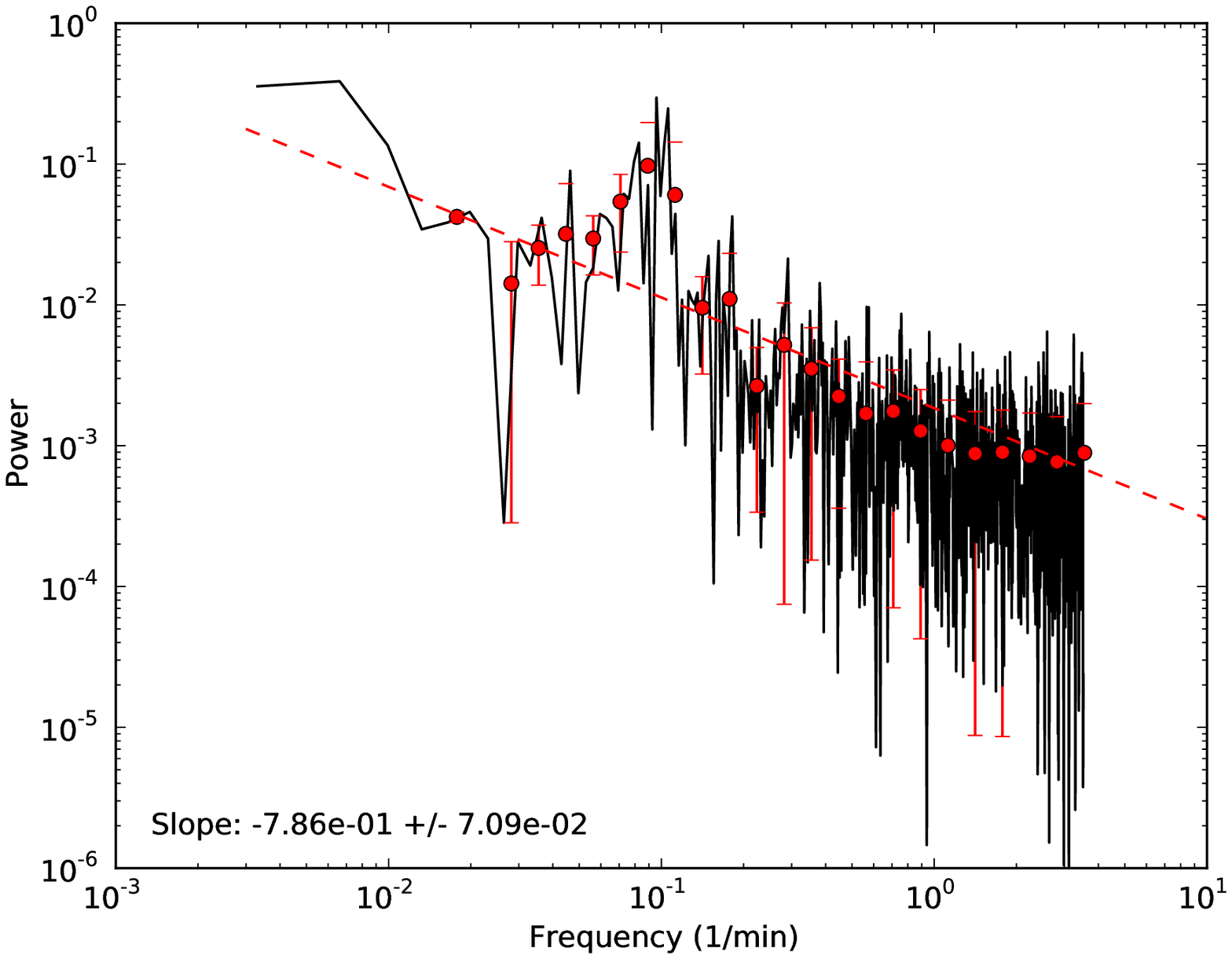}
\includegraphics[scale=0.35,angle=0]{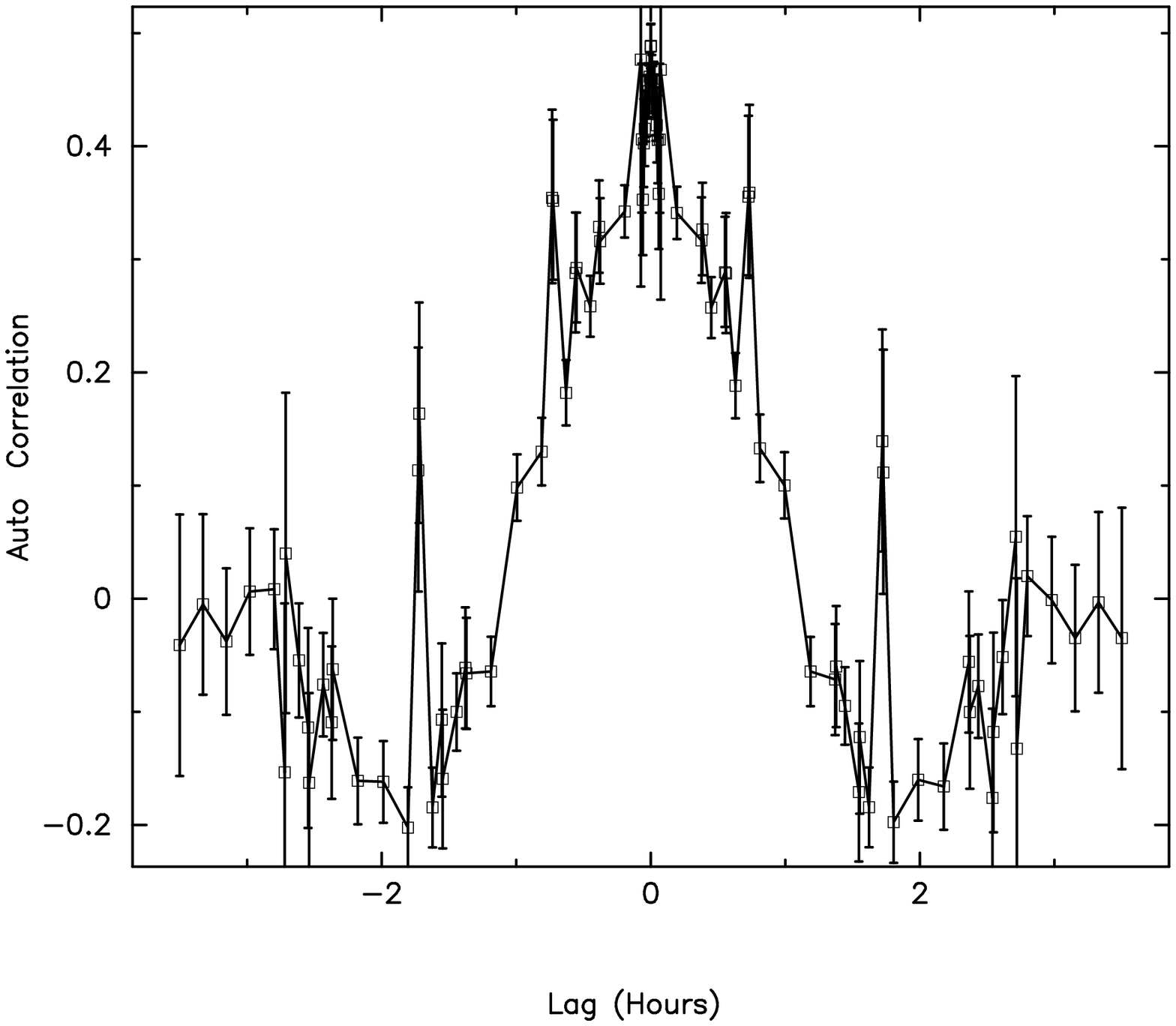}
\caption{
{\it (a) Top Left}
A structure function plot (units Jy$^2$) for VLBA observation of
Sgr~A* at 7mm on 2008, May 5 with data sampling of 3.3 sec.
The mean error is 0.050 Jy.
{\it (b) Top Right}
The CLEAN PSD of data shown in (a).
{\it (c) Bottom Left} The PSD of VLBA data similar  
that of Figure 2c. 
{\it (d) Bottom Right}
A plot of the autocorrelation function for VLBA data. 
Power-law fits to the structure function   over a 
range between 1 and 10 minutes as well as to the corresponding PSD, as shown in  (c), 
over the entire frequencies are displayed on each  figure.  
}
\end{figure}

%Similar to (a) except at 13mm. VLA and VLBA observations
%were made simultaneously at 7mm.

\begin{figure}
\center
\includegraphics[scale=0.35,angle=0]{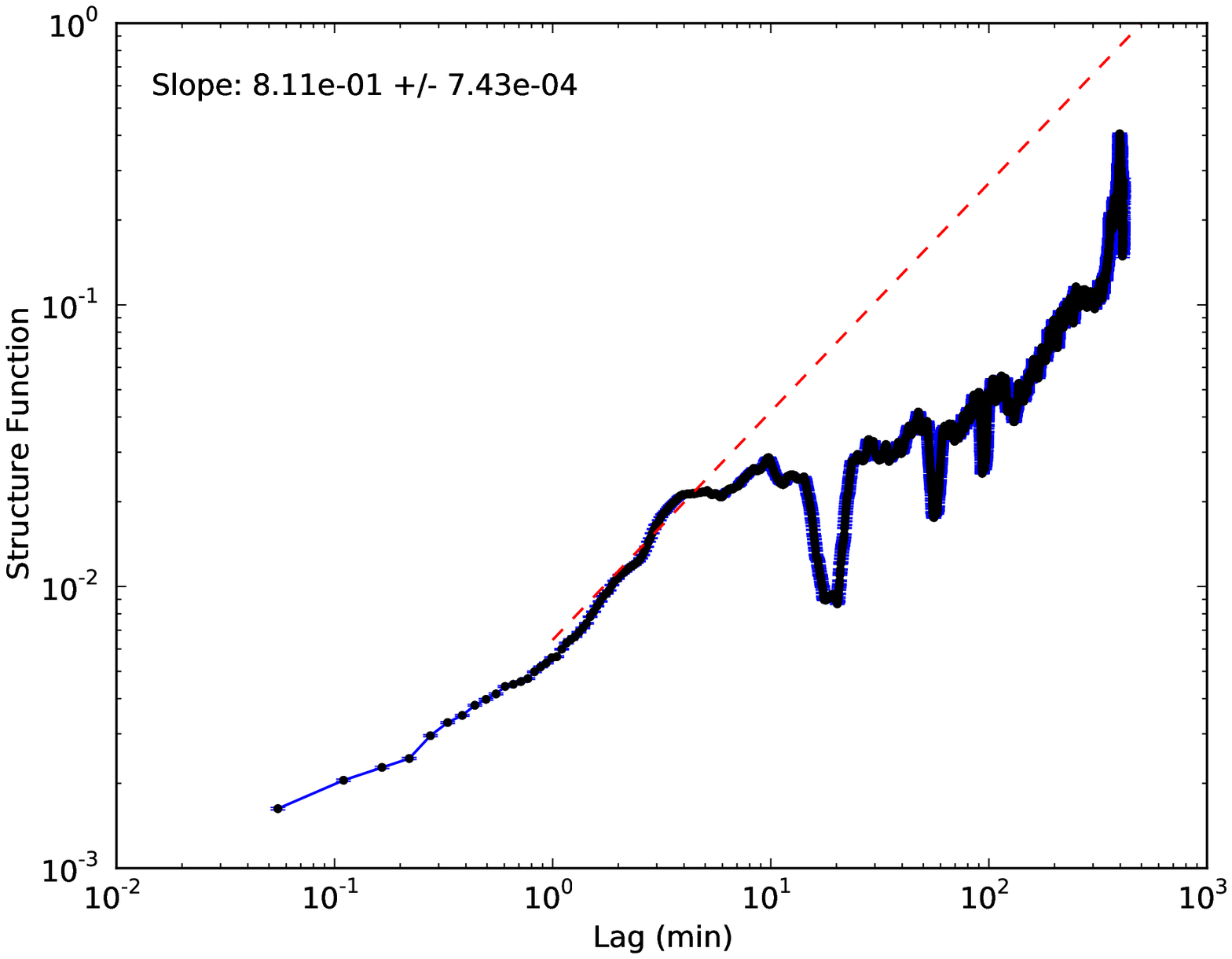}
\includegraphics[scale=0.35,angle=0]{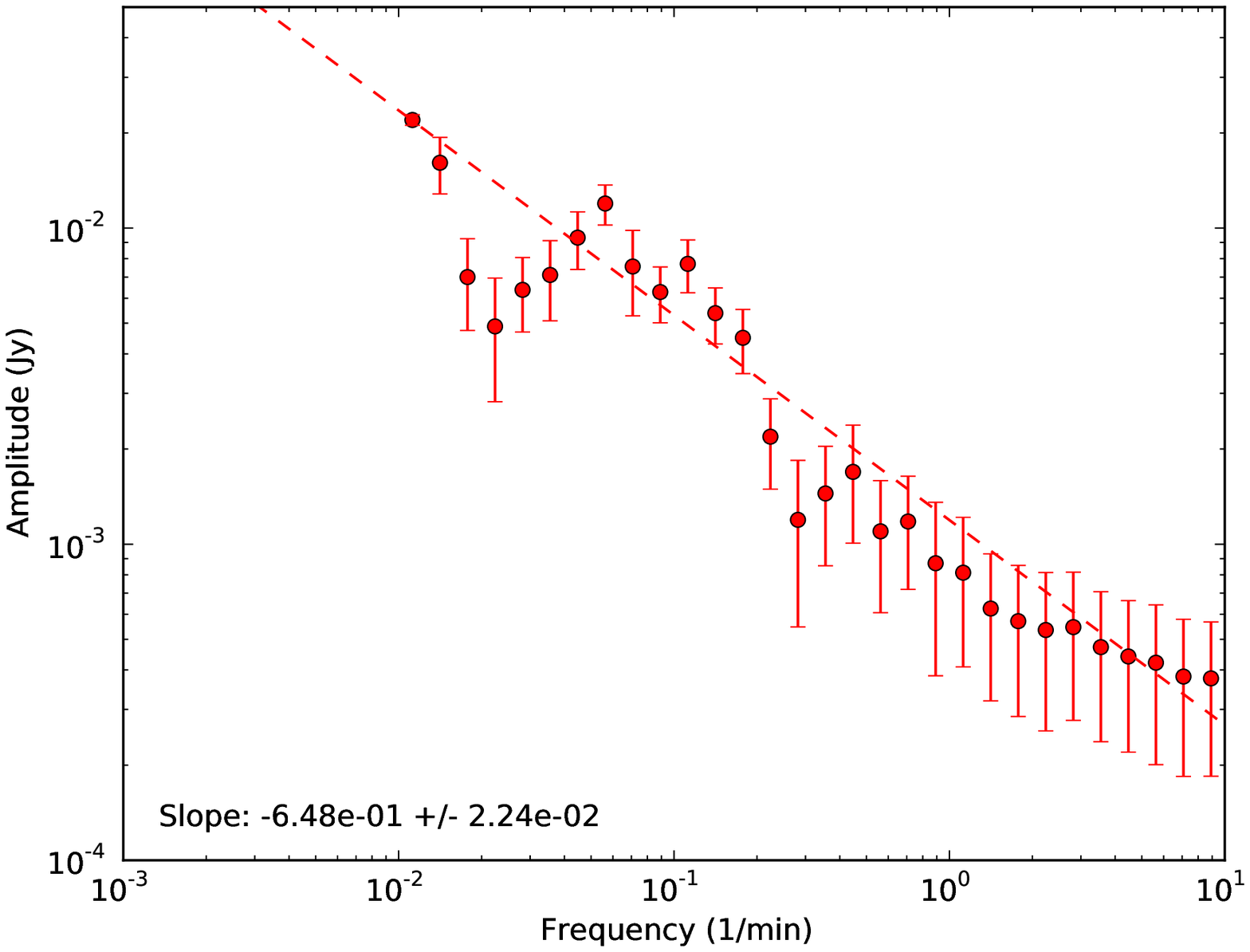}
\includegraphics[scale=0.35,angle=0]{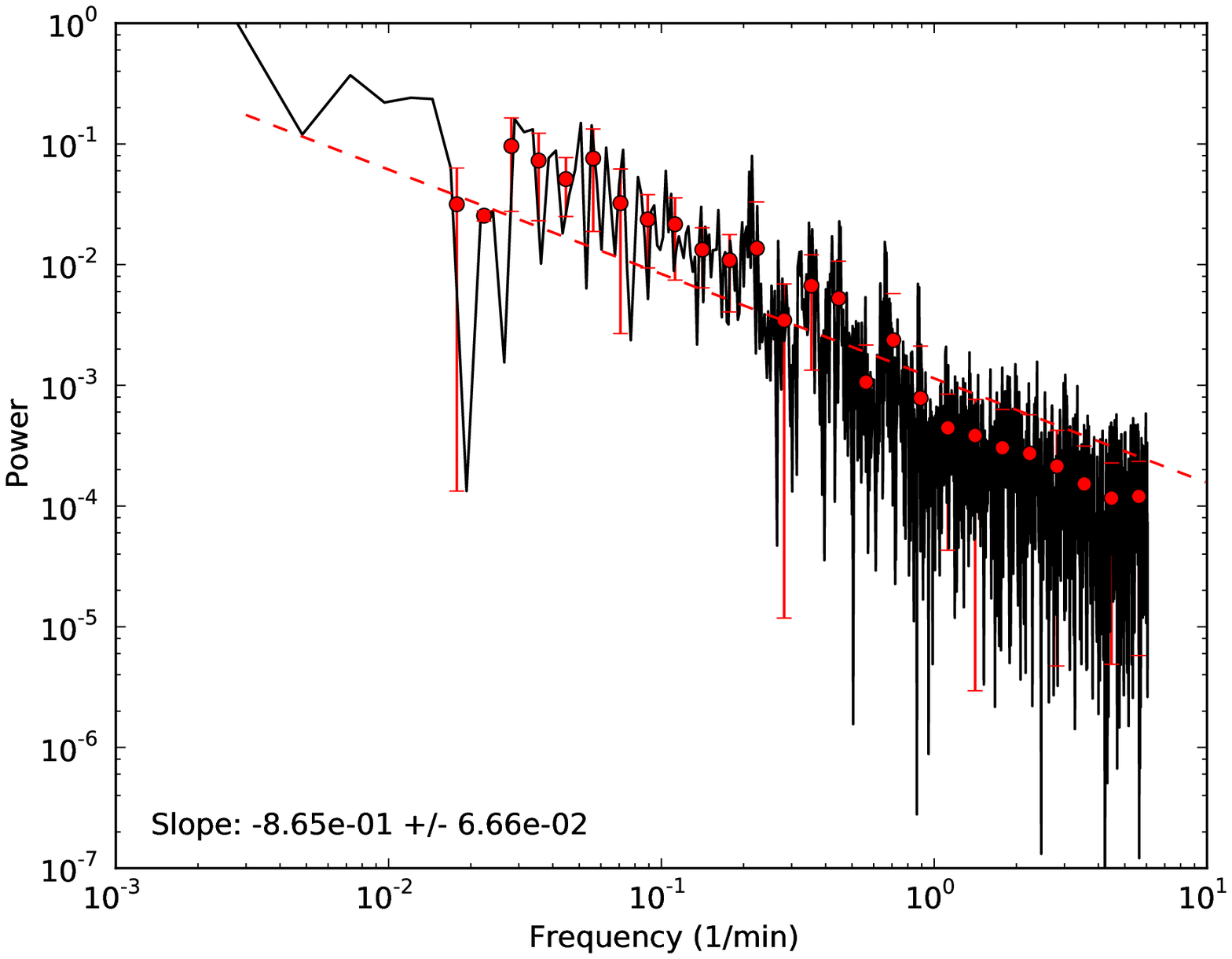}
\includegraphics[scale=0.35,angle=0]{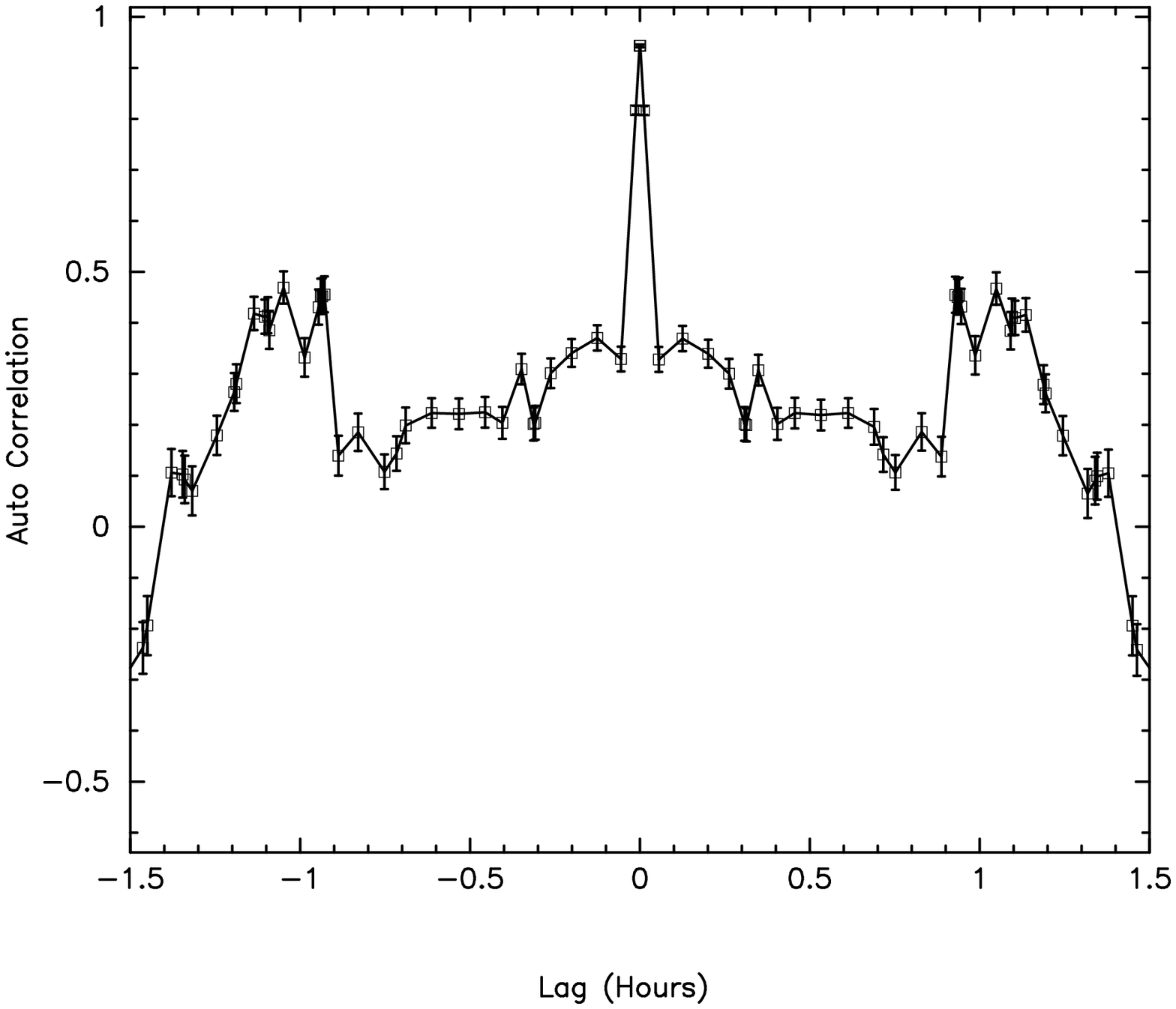}
\caption{
{\it (a) Top Left}
A structure function plot for VLA observation of
Sgr~A* at 7mm on 2008, May 10 with data sampling of 3.3 sec.
The mean error of individual data points  is 0.027 Jy.
The squares of the mean measurement
errors are 7.3$\times10^{-4}$ Jy$^2$ at 7mm. 
{\it (b) Top Right}
The corresponding CLEAN PSD of data shown in (a). The red dots represent  smoothed bins. 
{\it (c) Bottom Left} The  PSD using Uttley's technique with red dots show the binned data. 
The slope fitted through the entire spectrum is displayed. 
{\it (d) Bottom Right} 
A plot of the autocorrelation function. 
Power-law fits to the structure function   over a 
range between 0.1 and 1 minutes as well as to the corresponding power spectra 
over the entire frequencies are displayed on each  figure.  
}
\end{figure}

\begin{figure}
\center
\includegraphics[scale=0.35,angle=0]{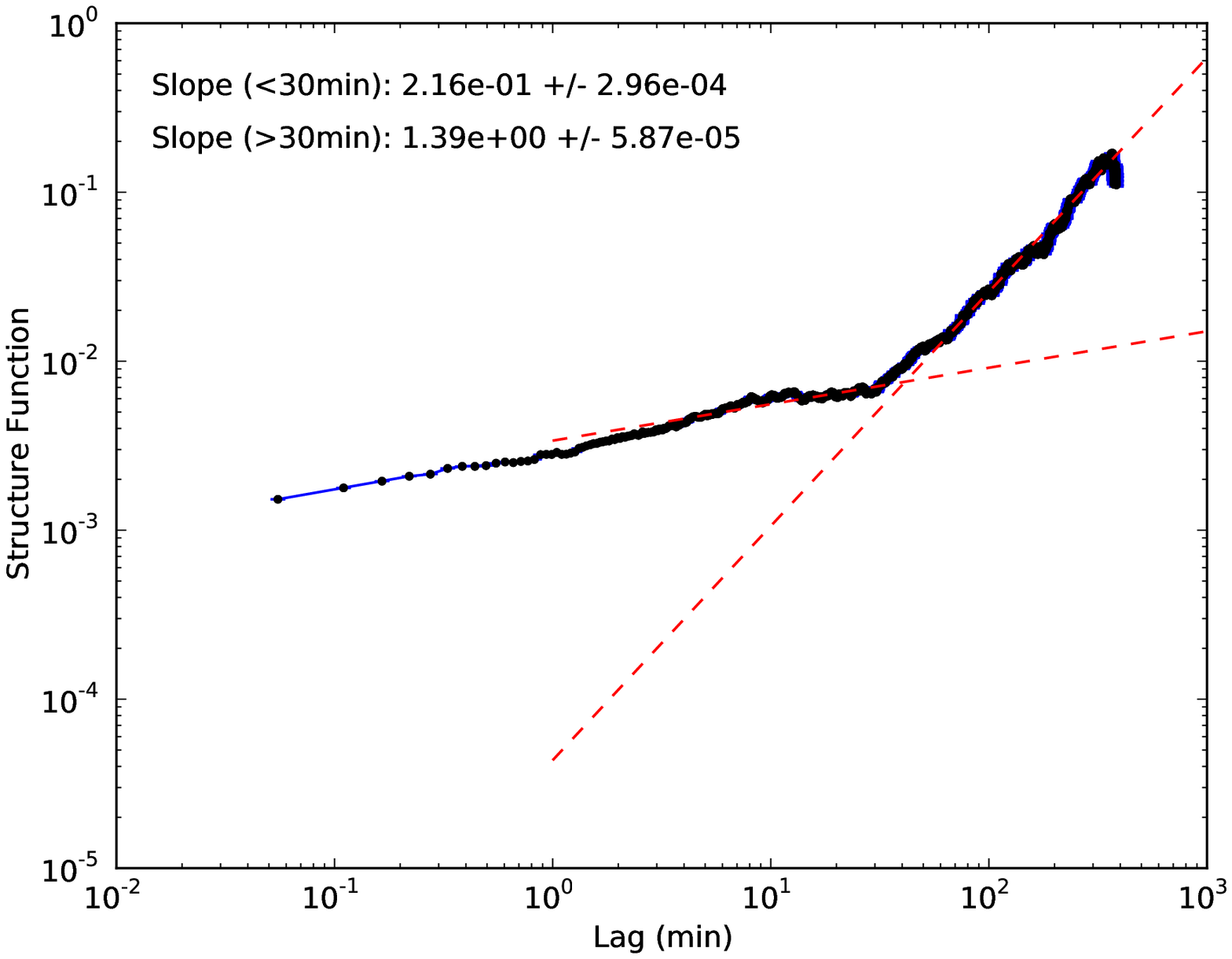}
\includegraphics[scale=0.35,angle=0]{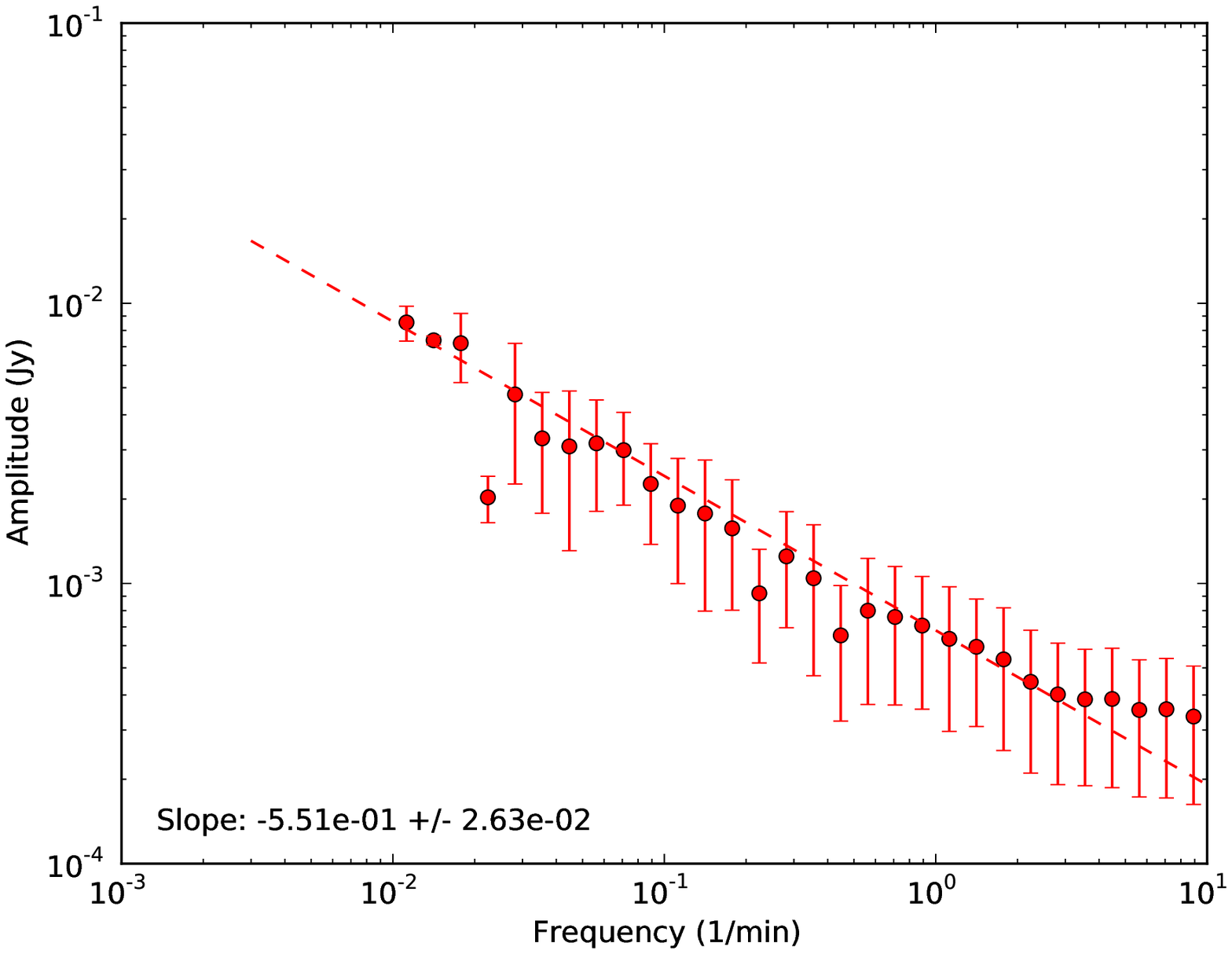}
\includegraphics[scale=0.35,angle=0]{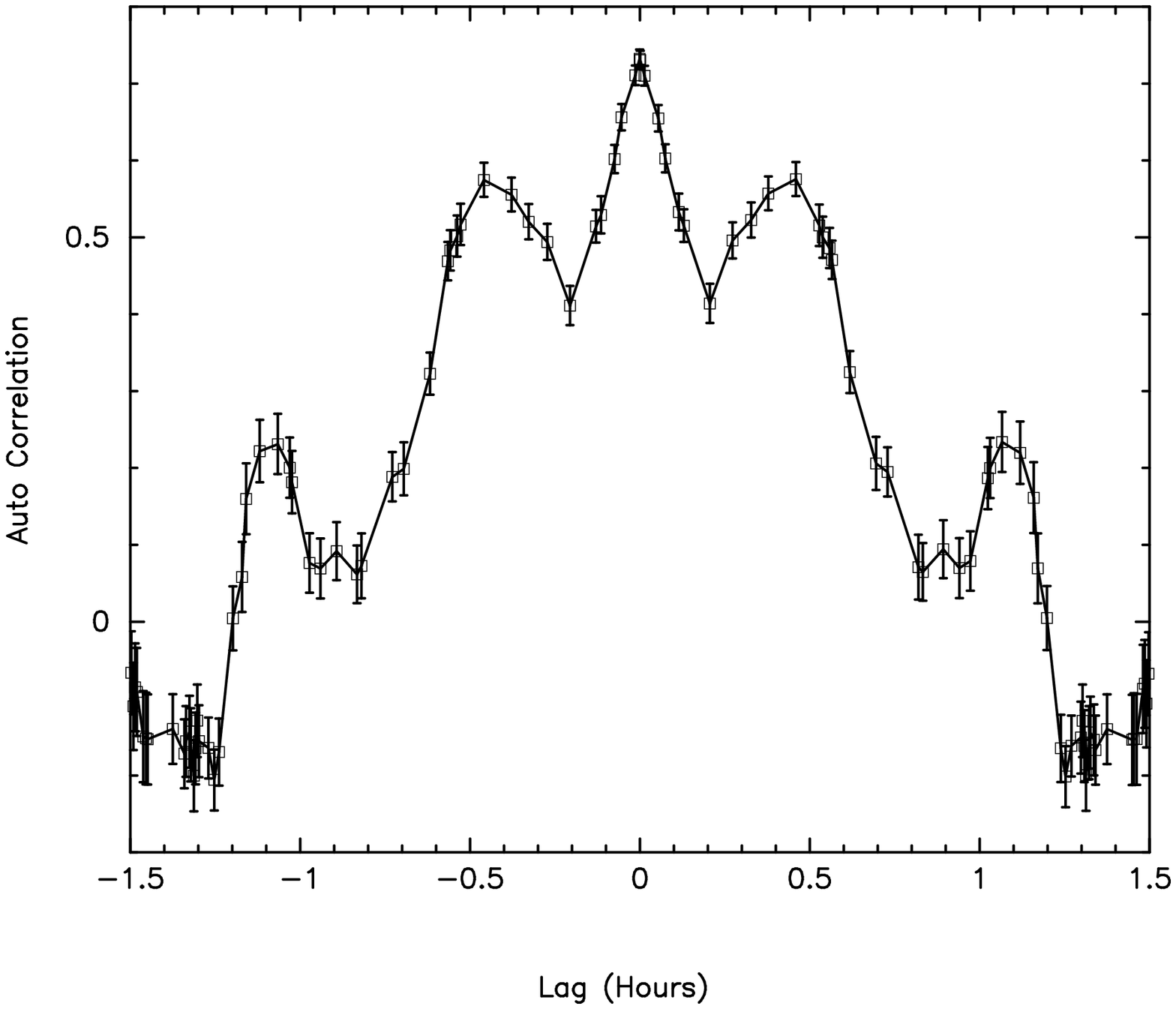}
\caption{
{\it (a) Top Left}
A structure function plot for VLA observation of
Sgr~A* at 7mm on 2008, May 6 with data sampling of 3.3 sec.
The mean error is 0.026 Jy.
The squares of the mean measurement
errors are 6.9$\times10^{-4}$ Jy$^2$ at 7mm.
{\it (b) Top Right}
The corresponding CLEAN PSD of data shown in (a). The red dots represent  smoothed bins. 
{\it (c) Bottom} 
A plot of the autocorrelation function for data.
Power-law fits to the structure function   over a 
range between 1 and 10 minutes as well as to the corresponding power spectrum 
over the entire frequencies are displayed on each  figure.  
}
\end{figure}

%(c - Bottom Left)
%The red dots show binned data when averaged for  a given lagtime.
%The slope to all  plots are presented as dashed lines.
%(d - Bottom Right)
%The corresponding power spectral density of (b).

\begin{figure}
\center
\includegraphics[scale=0.35,angle=0]{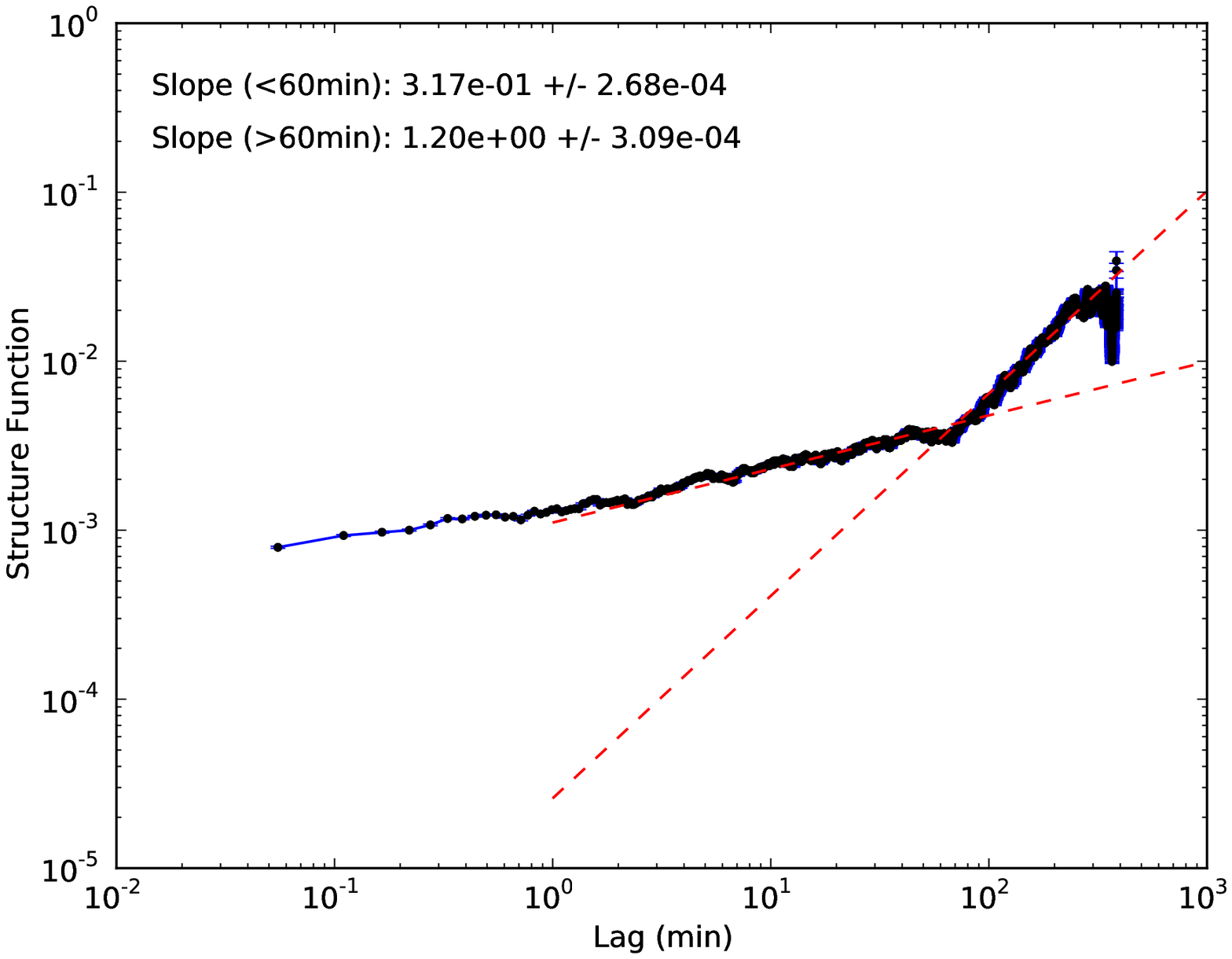}
\includegraphics[scale=0.35,angle=0]{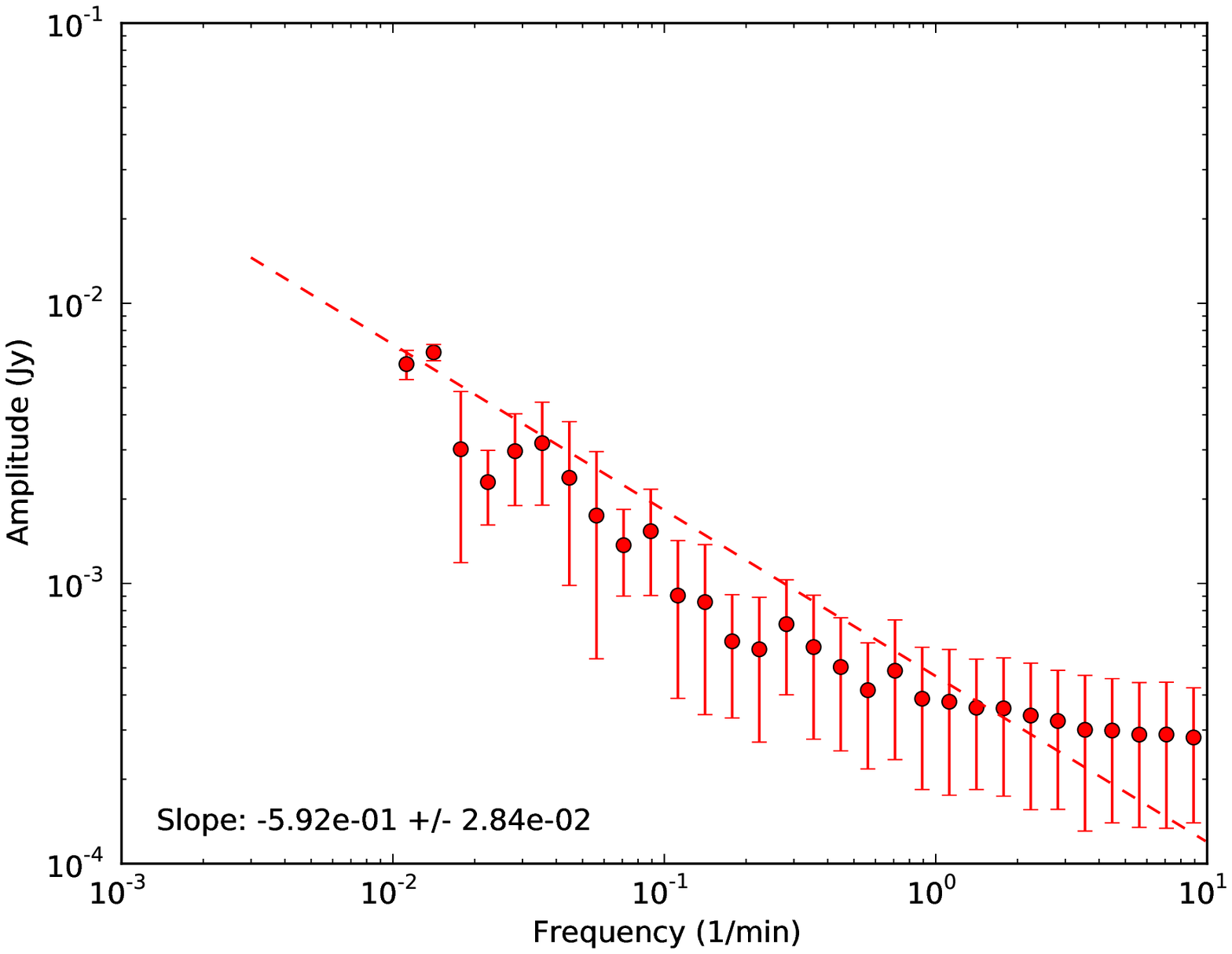}
\includegraphics[scale=0.35,angle=0]{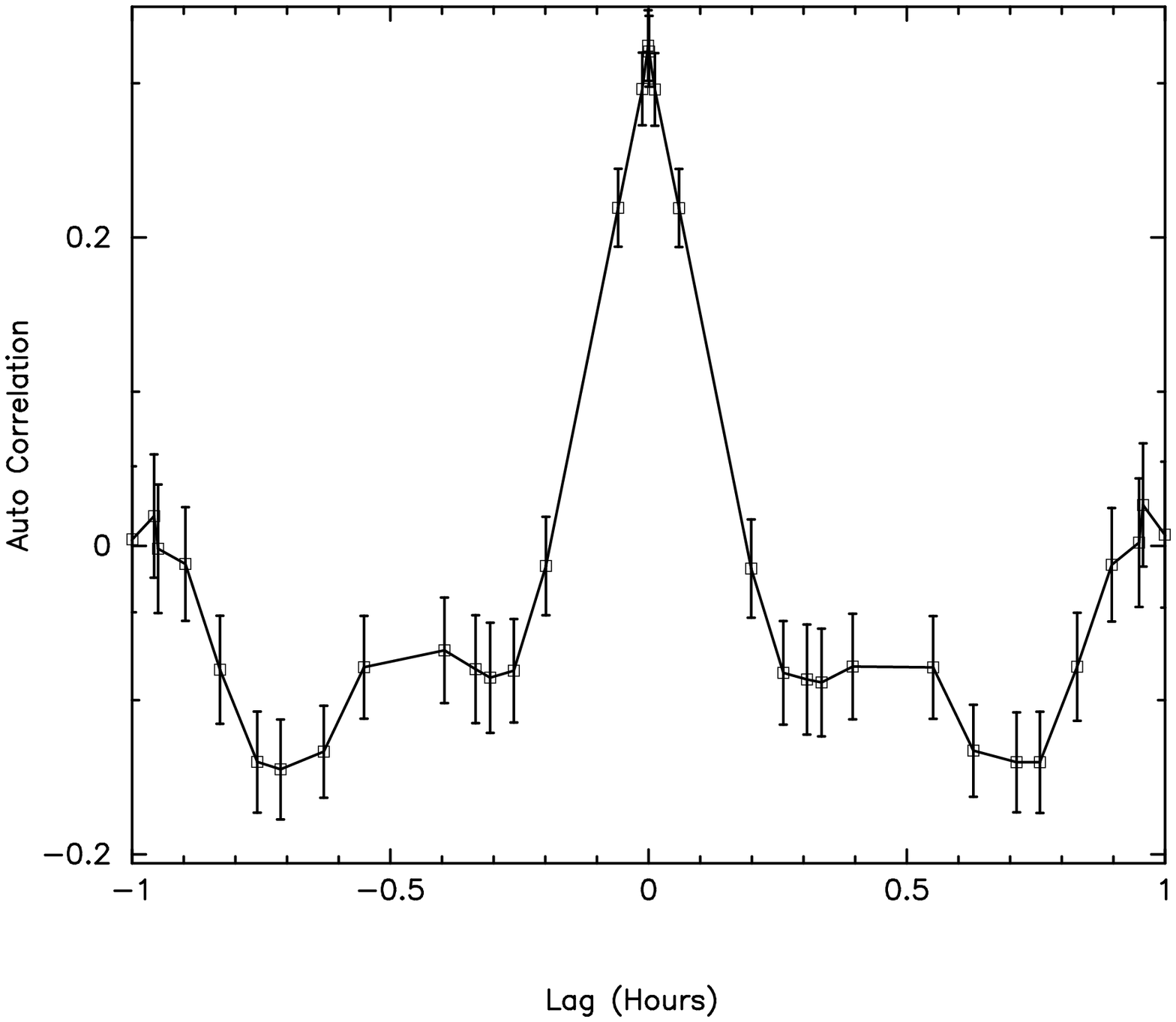}
\caption{
{\it (a) Top Left}
A structure function plot for VLA observation of
Sgr~A* at 13mm on 2008, May 6 with data sampling of 3.3 sec.
The mean error is 0.03 Jy and the the square of the mean measurements
errors are 9$\times10^{-4}$ Jy$^2$ at 13mm.
{\it (b) Top Right}
The CLEAN PSD of data shown in (a). The red dots represent  smoothed bins. 
{\it (c) Bottom}
A plot of the autocorrelation function for data. 
Power-law fits to the structure function   over a 
range between 1 and 100 minutes as well as to the corresponding power spectra 
over the entire frequencies are displayed on each  figure.  
}
\end{figure}

\begin{figure}
\center
\includegraphics[scale=0.35,angle=0]{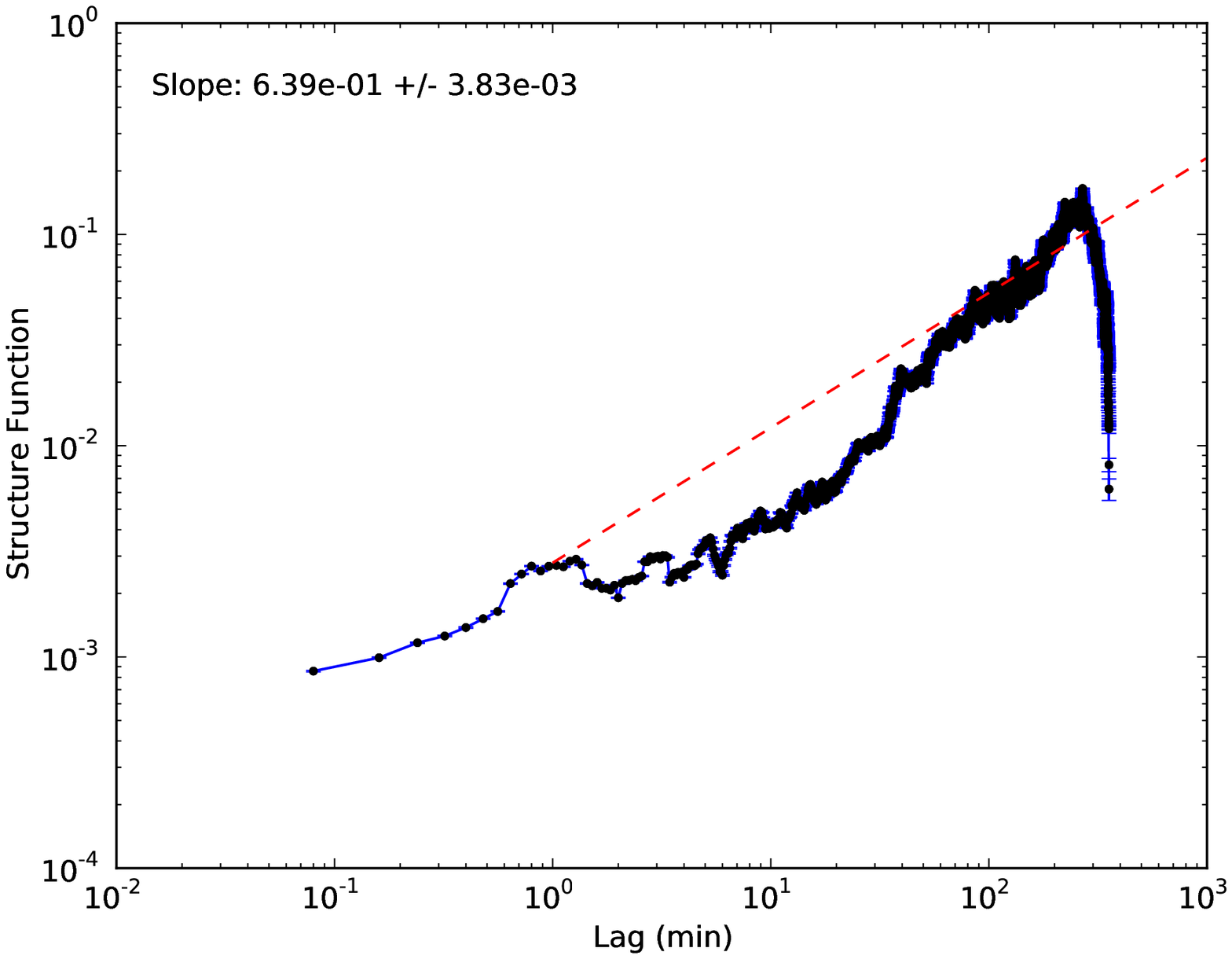}
\includegraphics[scale=0.35,angle=0]{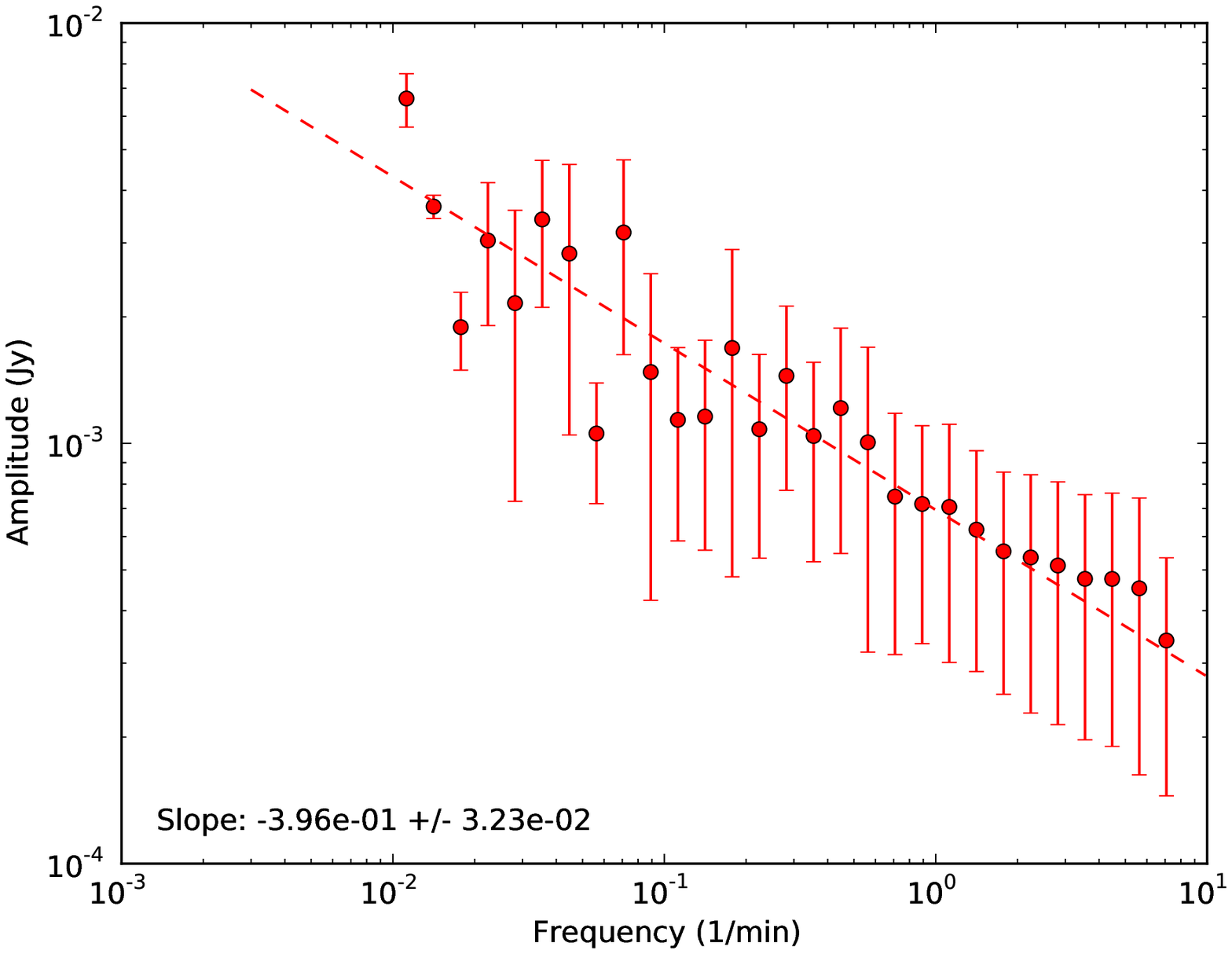}
\includegraphics[scale=0.35,angle=0]{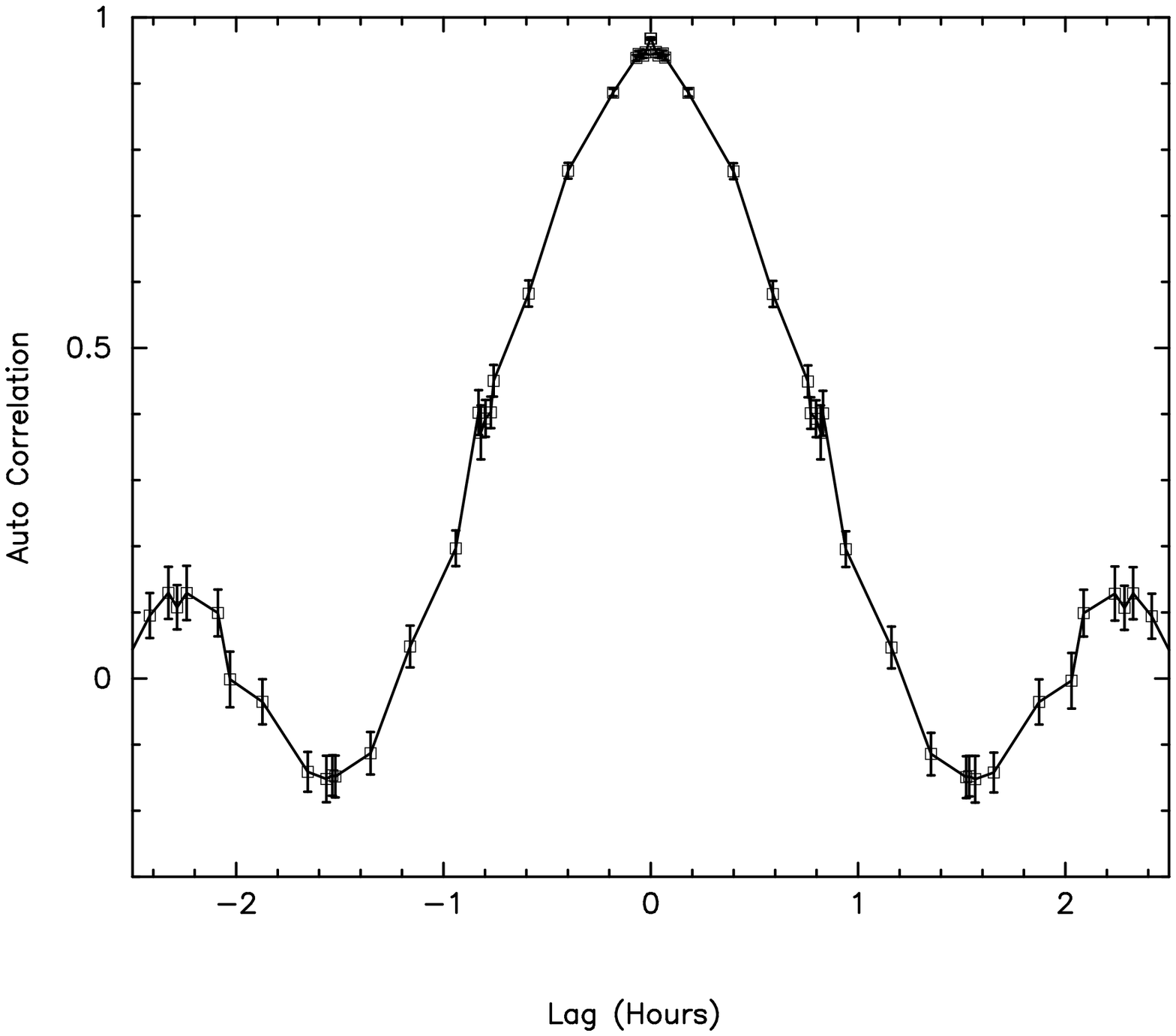}
\caption{
{\it (a) Left}
A structure function plot for VLA observation of
Sgr~A* at 7mm on 2006, Feb 10 with data sampling of 3.3sec.
The power law fit is shown in dashed red lines.
The amplitude are in  Jy$^2$.
The mean errors is 0.009 Jy.
The squares of the mean measurement
errors are 0.9$\times10^{-4}$ Jy$^2$ at 7mm.
{\it (b) Top Right}
The CLEAN PSD of data shown in (a). The red dots represent  smoothed bins. 
{\it (c) Bottom} 
A plot of the autocorrelation function. 
Power-law fits to the structure function   over a 
range between 0.1 and 1 minutes as well as to the corresponding power spectra 
over the entire frequencies are displayed on each  figure.  
}
\end{figure}

\begin{figure}
\center
\includegraphics[scale=0.35,angle=0]{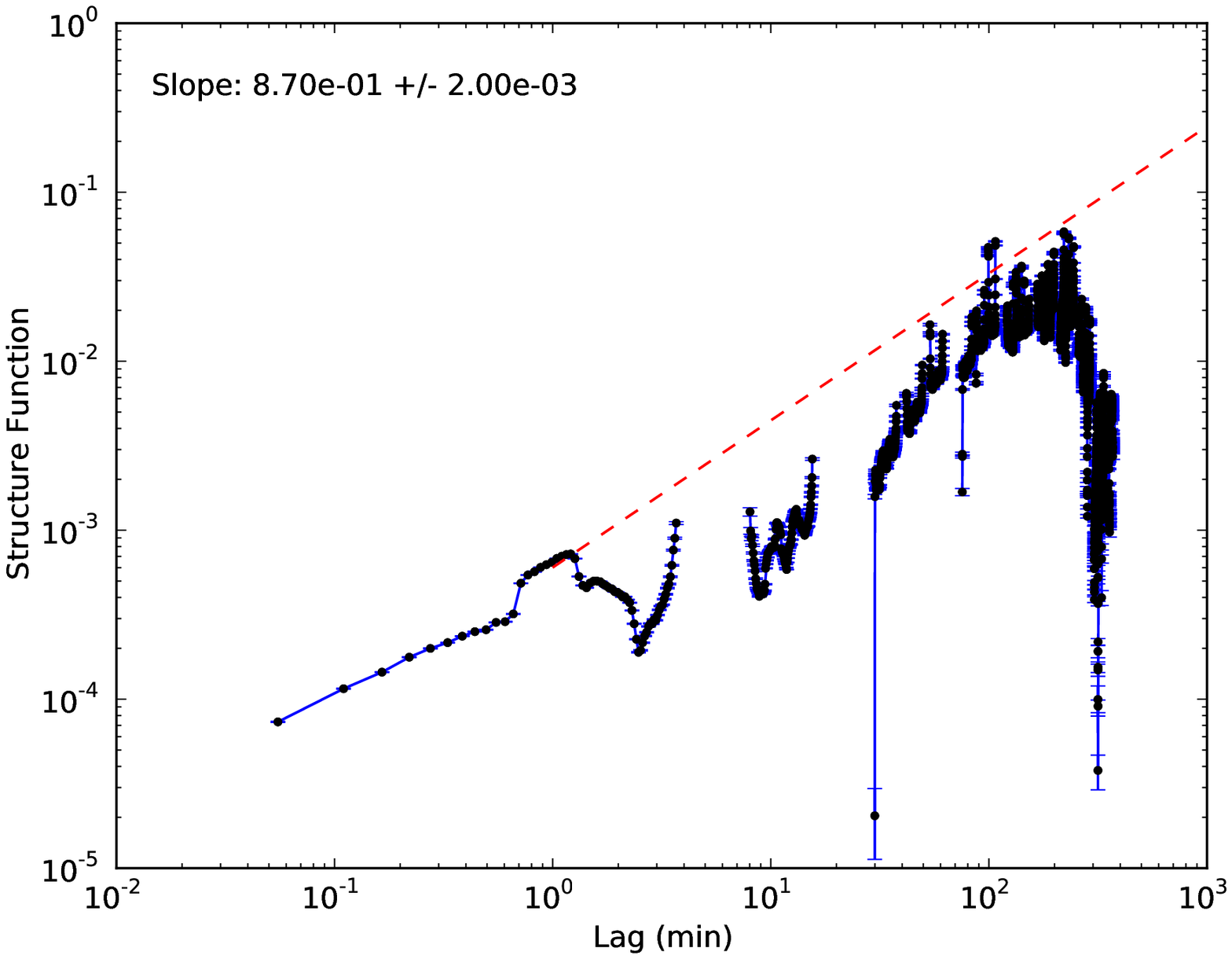}
\includegraphics[scale=0.35,angle=0]{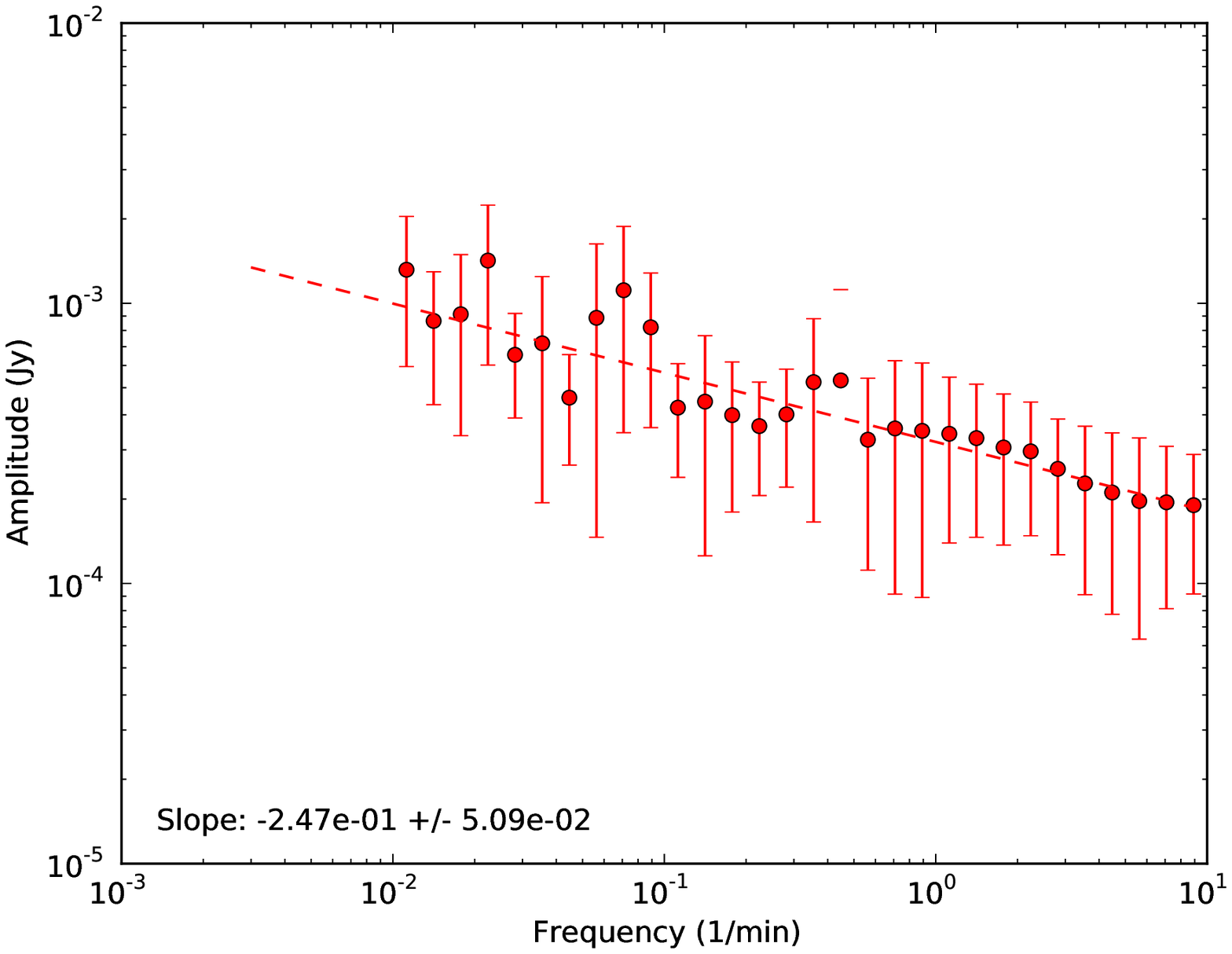}
\includegraphics[scale=0.35,angle=0]{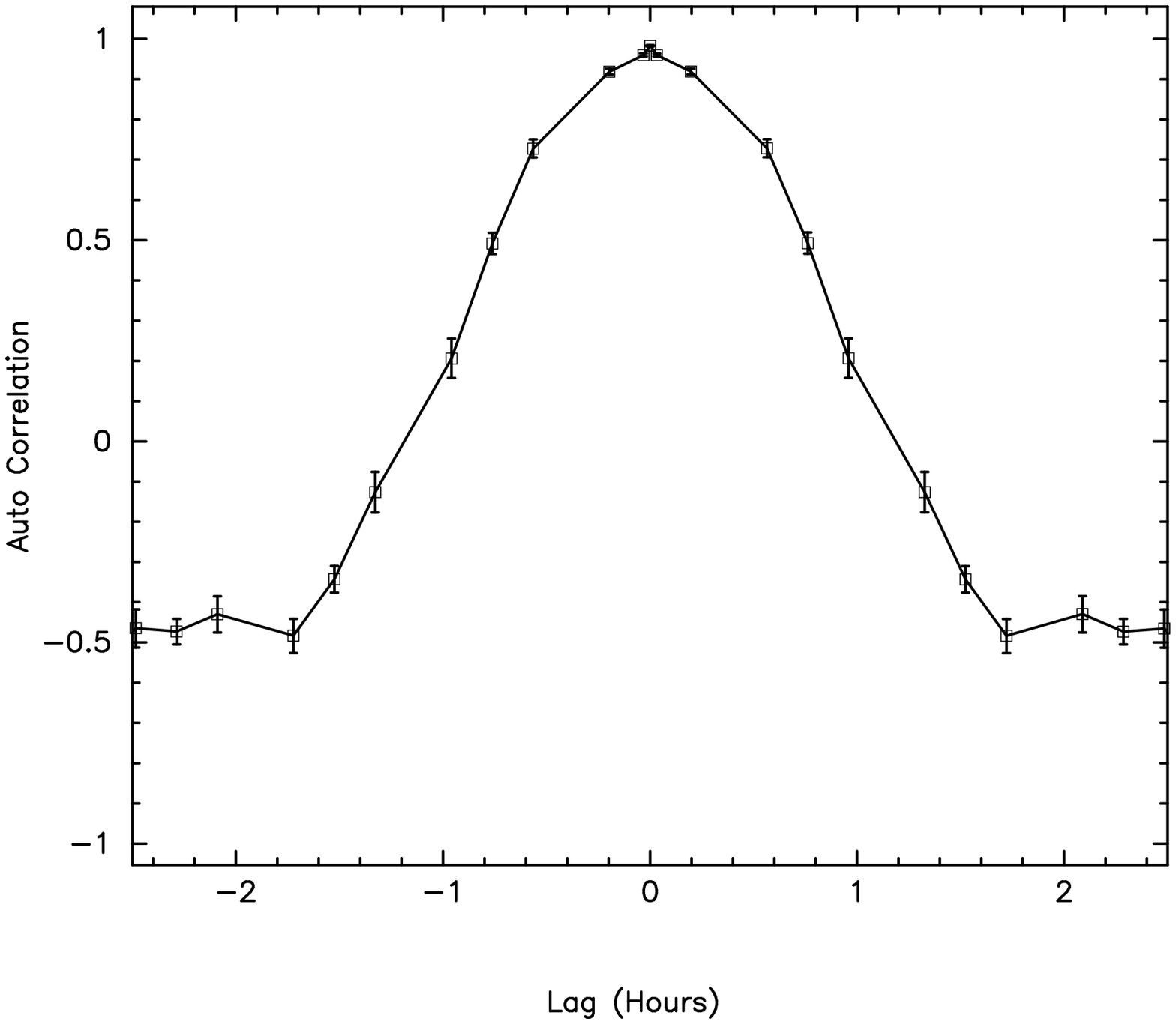}
\caption{
{\it (a) Top Left}
A structure function plot for VLA observation of
Sgr~A* at 13mm on 2006, Feb 10 with data sampling of 3.3 sec.
The power law fit is shown in dashed red lines.
The amplitude are in  Jy$^2$.
The mean errors is 0.003Jy.
The squares of the mean measurement
errors are 9$\times10^{-6}$ Jy$^2$ at 13mm.
{\it (b) Top Right}
The CLEAN PSD of data shown in (a). The red dots represent  smoothed bins. 
{\it (c) Bottom} 
A plot of the autocorrelation function. 
Power-law fits to the structure function   over a 
range between 0.1 and 1 minutes as well as to the corresponding power spectra 
over the entire frequencies are displayed on each  figure.  
}
\end{figure}

\begin{figure}
\center
\includegraphics[scale=0.35,angle=0]{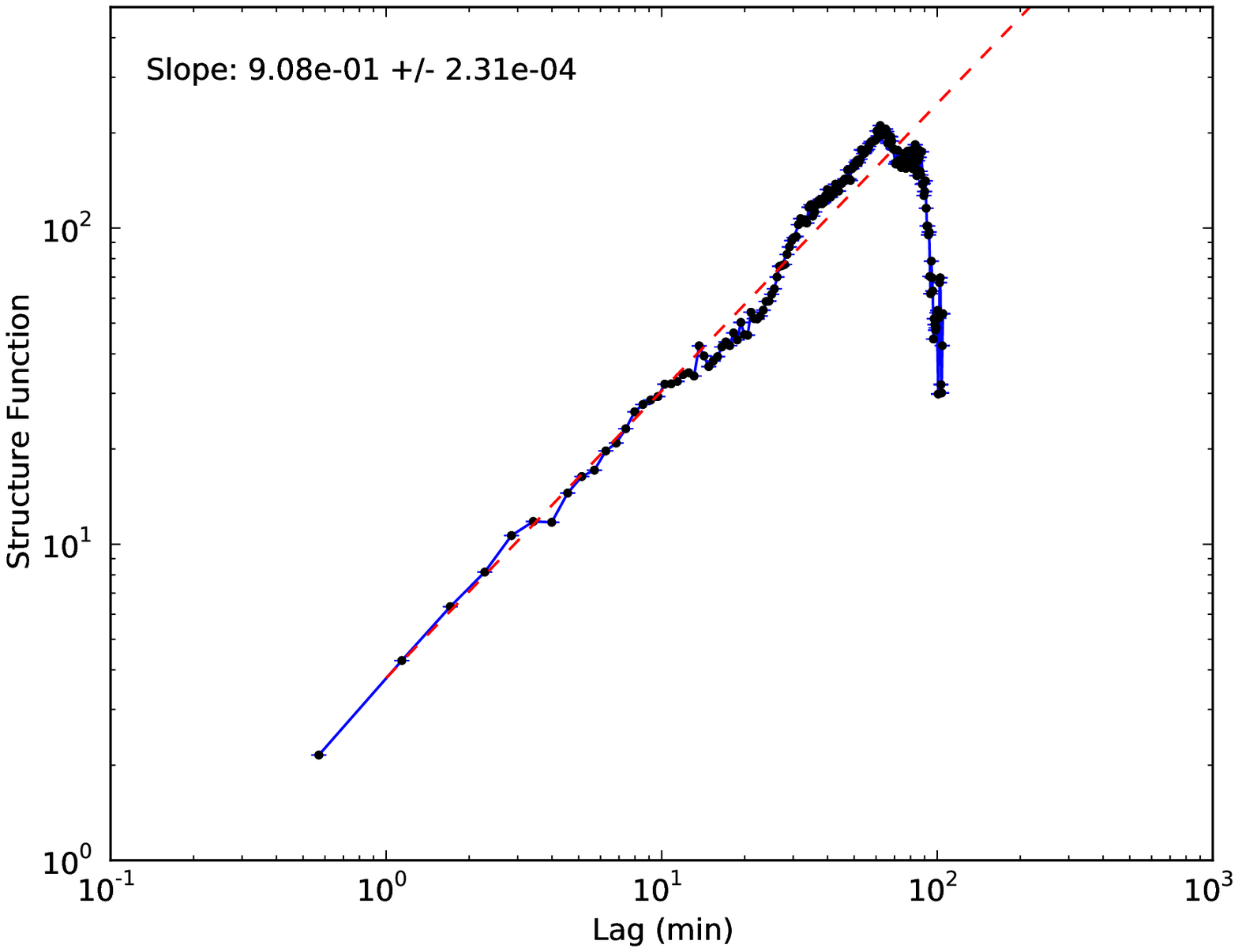}
\includegraphics[scale=0.35,angle=0]{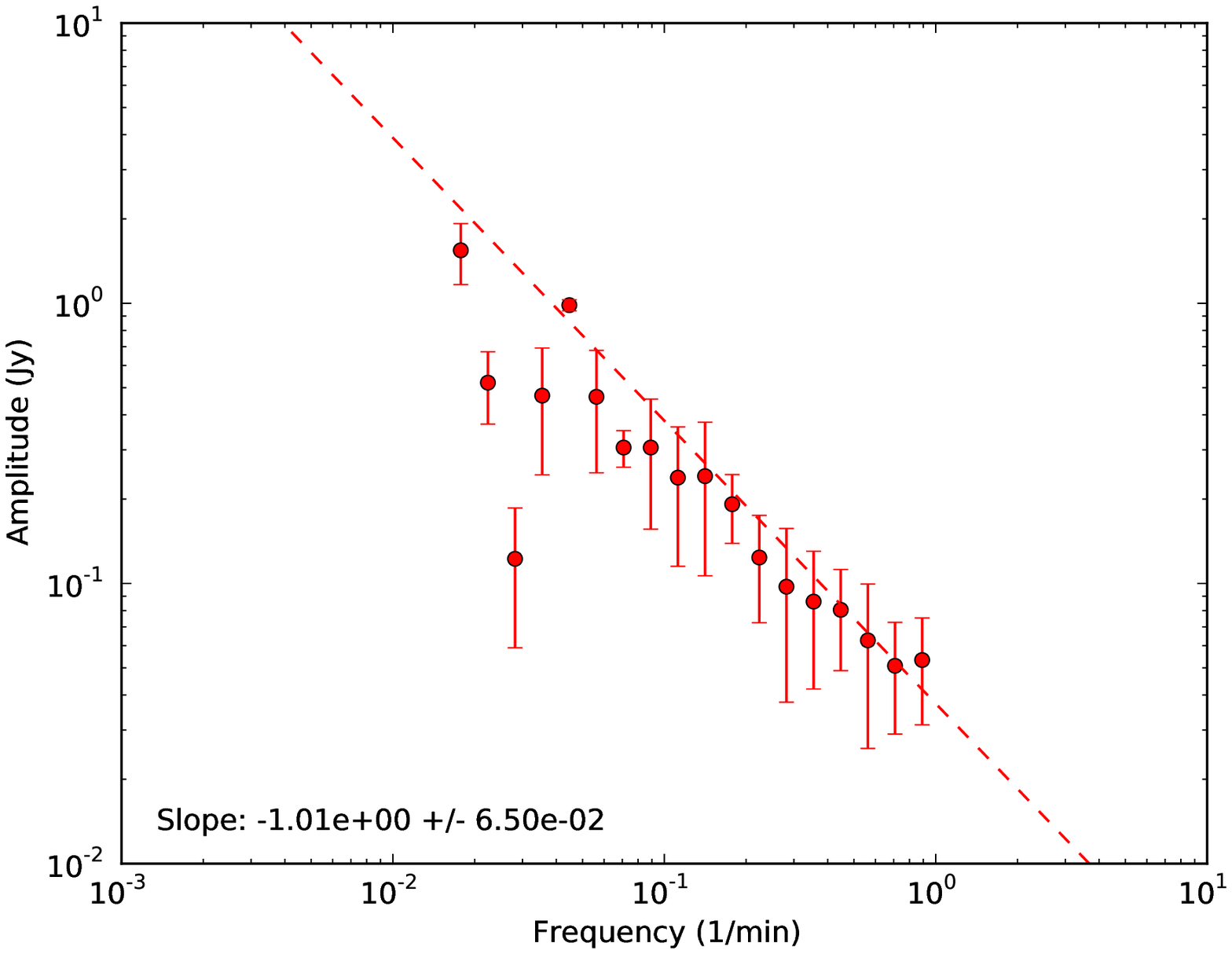}
\includegraphics[scale=0.35,angle=0]{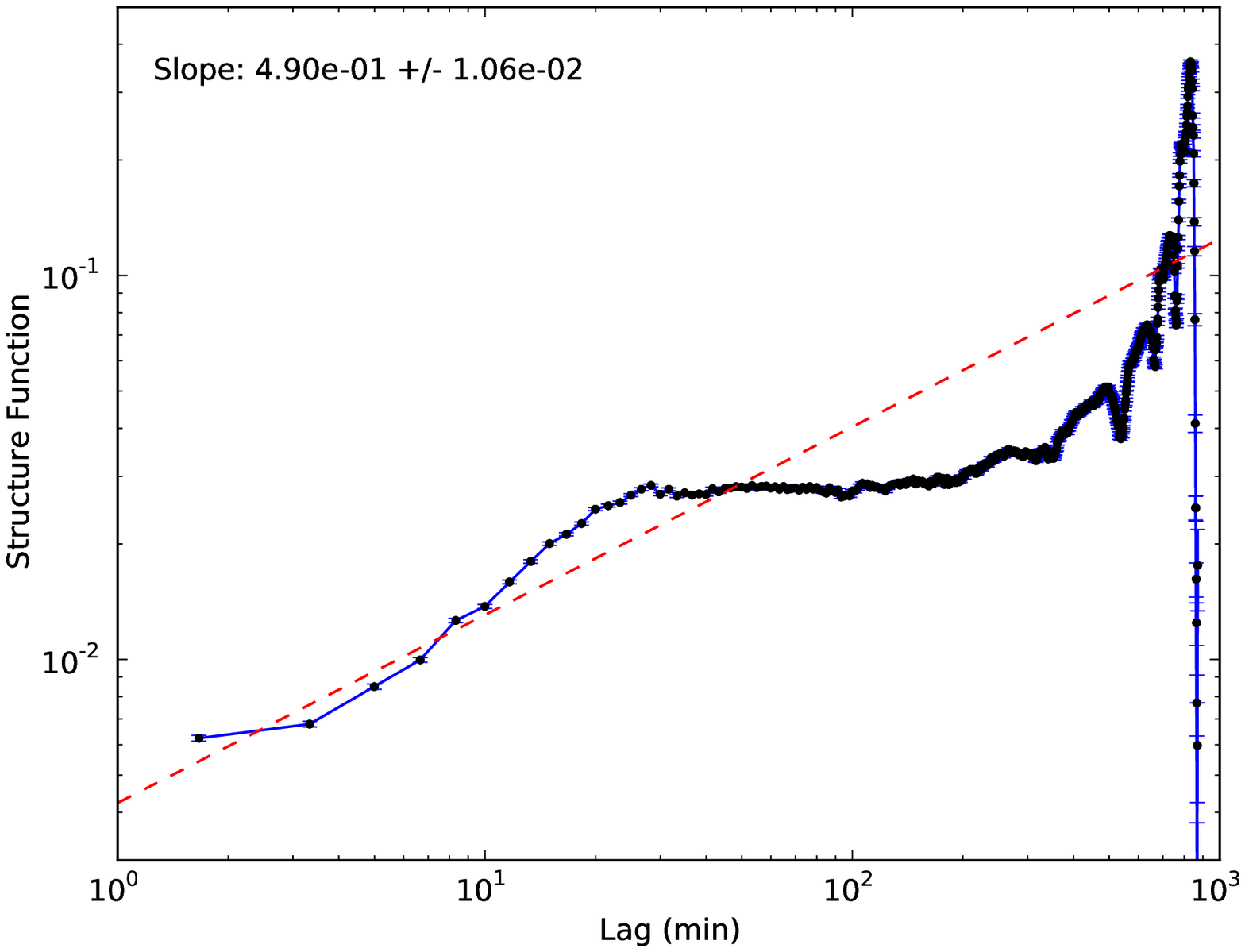}
\includegraphics[scale=0.35,angle=0]{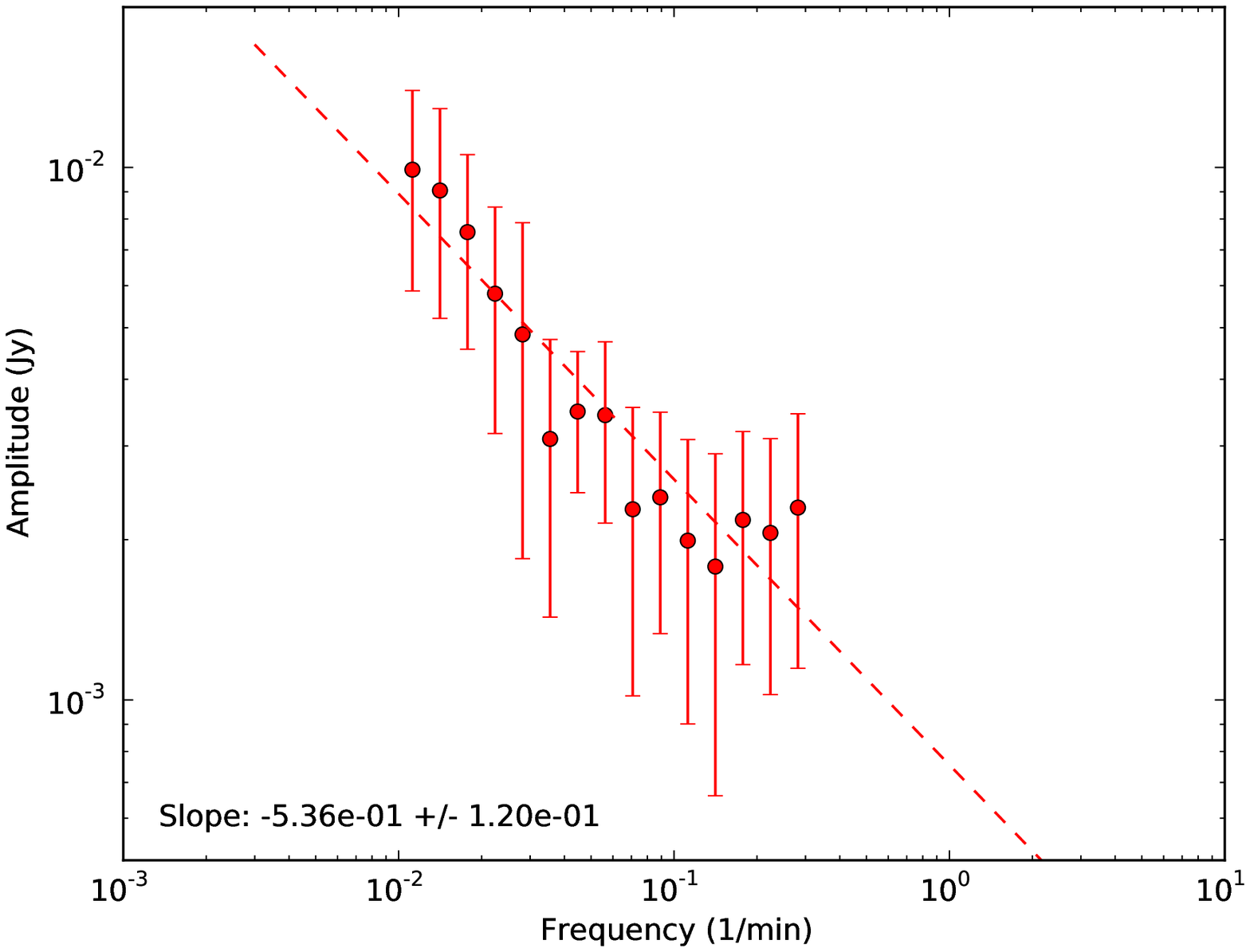}
\includegraphics[scale=0.35,angle=0]{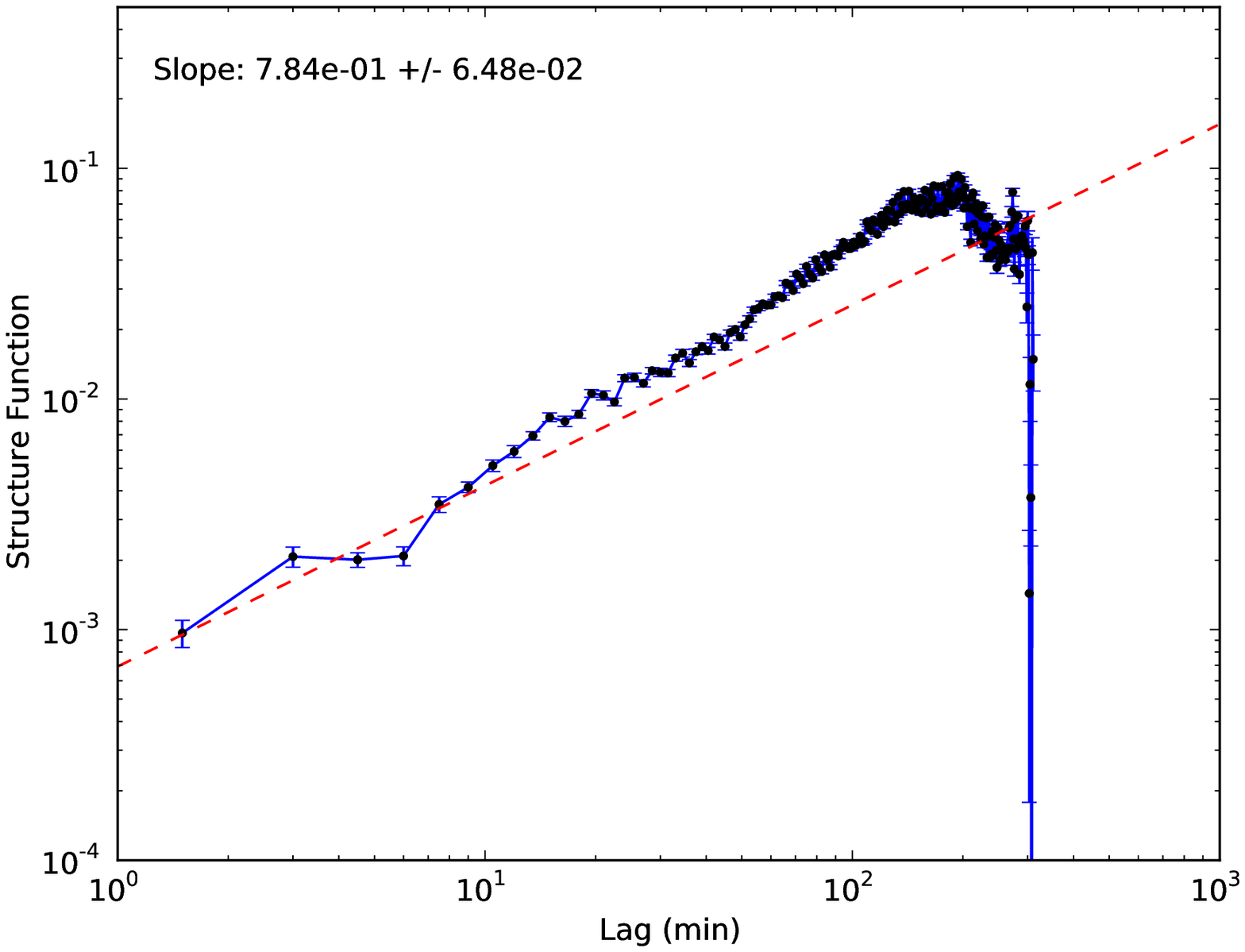}
\includegraphics[scale=0.35,angle=0]{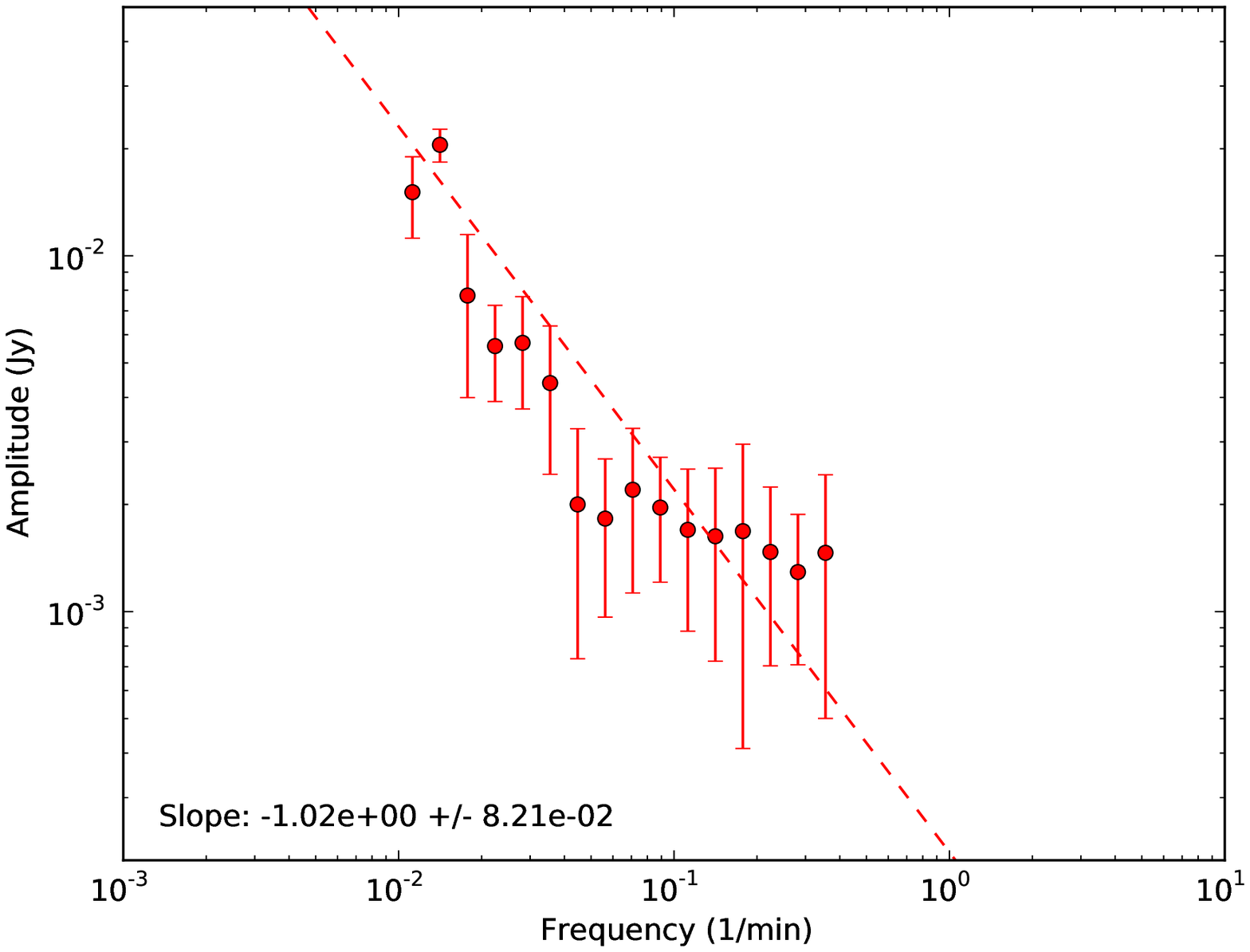}
\caption{
{\it (a) Top Left}
A structure function plot for a day of observation
on 2007, April 4 using VLT at 3.8$\mu$m (Dodds-Eden et al. 2009).
The power law fit is shown by a dashed line.
{\it (b) Top Right}
Similar to (a) except the corresponding CLEAN PSD is shown. 
{\it (c) Middle  Left}
Similar to (a) except the data were taken with XMM-Newton
on the same day but longer observation.
There was considerable activity on this day in both IR and X-ray wavelengths
(Porquet et al. 2008).
{\it (d) Middle Right}
Similar to (b) showing the CLEAN PSD of X-ray data.
{\it (e) Bottom  Left}
Similar to (c) except that the data taken with the VLA at 7mm (Yusef-Zadeh et al. 2009).
The sampling time is 90 sec and the {\it uv} range is selected to be $> 100$
k$\lambda$.
{\it (f) Bottom Right}
Similar to (e) showing the PSD of radio data at 7mm. }
\end{figure}

\begin{figure}
\center
\includegraphics[scale=0.35,angle=0]{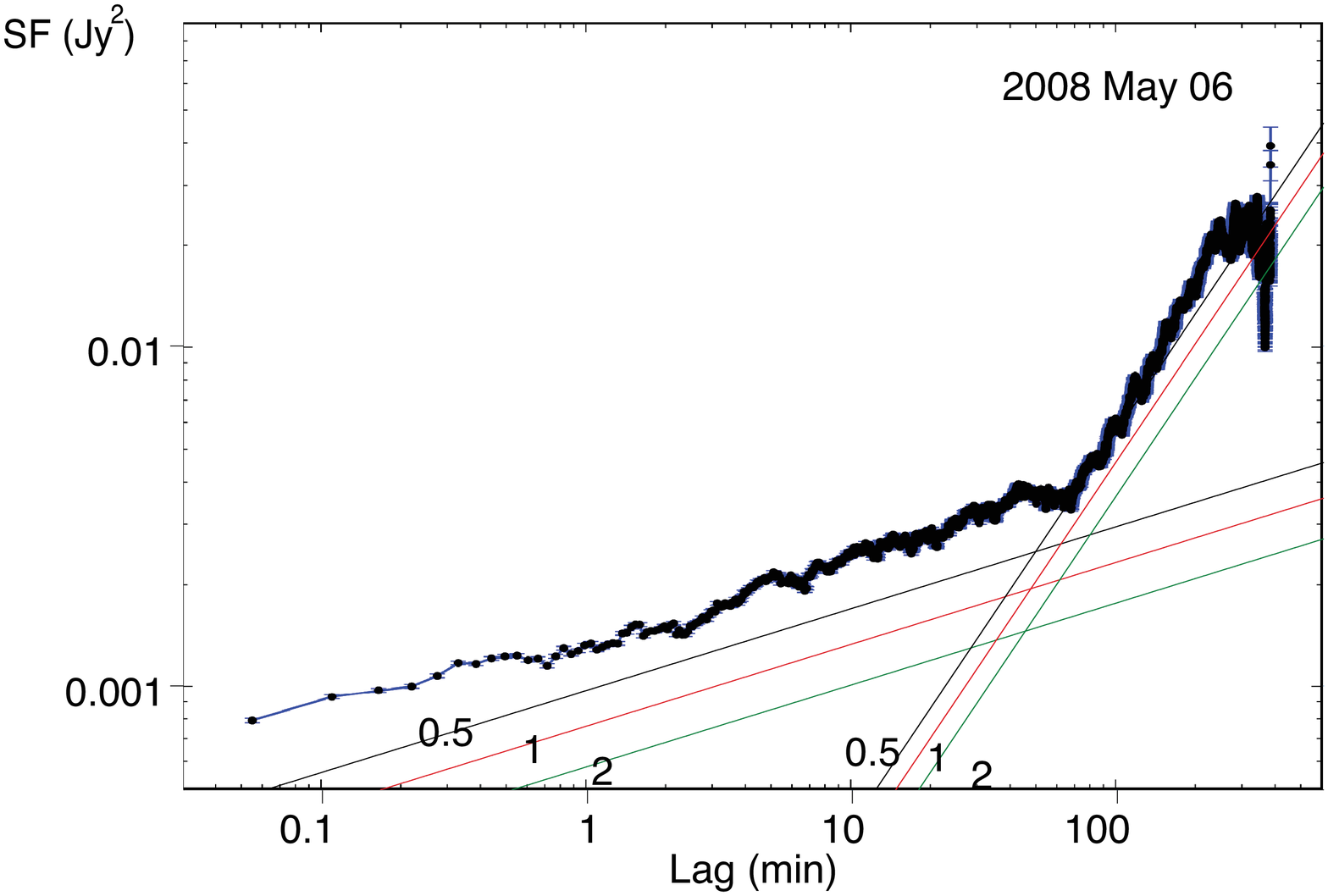}
\includegraphics[scale=0.35,angle=0]{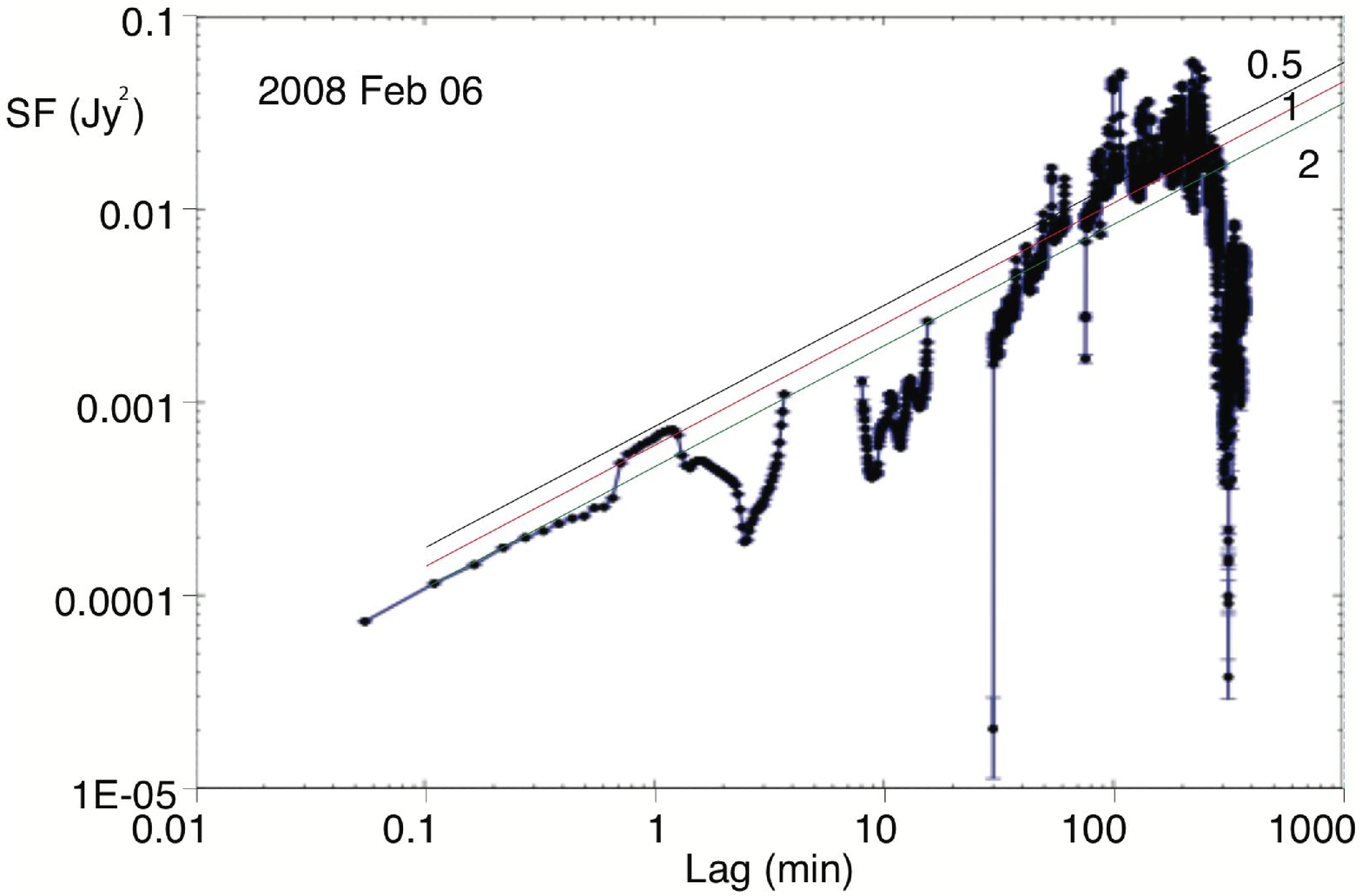}
\caption{
{\it (a) Top}
The 13mm structure function from Figure  6a with overlays of the prediction based on the
7mm structure function fits,  assuming power-law particle energy spectral indices p=0.5, 1, and 2.
Both power law components are well reproduced, so that the comparison of the SF at 7 and 13mm is
consistent with the plasmon model prediction.
{\it (b) Bottom}
Similar to (a) except  for the 2006 February 10 structure functions, as shown in Figure 8a.
}
\end{figure}


\begin{thebibliography}{99}
\bibitem[Alexander(1997)]{alexander97} Alexander, T. 1997,  \mnras\,  285, 891
\bibitem[Arshakian et al.(2010)]{arshakian10} Arshakian, T.~G., Leon-Tavares, J., Lobanov, A. P. et al. 2010, \mnras, 401, 1231
\bibitem[Barth et al.(2003)]{barth03} Barth, A. J., Ho, J. C., \& Sargent, W. L. W. 2003, \apj, 583, 134
\bibitem[Bower et al.(2004)]{bower04} Bower, G. C., Falcke, H., Herrnsein, R. M., Zhao J.-H., Goss, W. M. \& Backer, D. C. 2004, Science, 304, 704
\bibitem[Chan et al.(2009)]{chan09} Chan, C.-K., Liu, S., Fryer, C.~L., Psaltis, D., Ozel, F., Rockefeller, G. and Melia, F. 2009, \apj, 701, 521
\bibitem[Do et al.(2009)]{do09} Do, T.,  Ghez, A. M.,  Morris, M. R.,  Yelda, S.,  Meyer, L.,  Lu, J. R., Hornstein, S. D. \&  Matthews, K. 2009, \apj, 691, 1021
\bibitem[Doeleman et al.(2008)]{doeleman08} Doeleman, S.~S., Weintroub, J., Rogers, A.~E.~E., Plambeck, R., Freund, R.,  Tilanus, R.~P.~J. et al.  2008, \nat, 455, 70
\bibitem[Dodds-Eden et al.(2009)]{dodds09} {Dodds-Eden}, K. et al. 2009, \apj, 698, 676
\bibitem[Eckart et al.(2006)]{eckart06} Eckart, A., Schodel, R.,  Meyer, L.,  Trippe, S. Ott, T. and Genzel, R. 2006, \aap, 455, 1
\bibitem[Edelson and Krolik (1988)]{edelson88} Edelson, R. A., \& Krolik, J. H. 1988, ApJ, 333, 646 
\bibitem[Emmanoulopoulos et al.2010]{emma10} Emmanoulopoulos, D., McHardy, I. M., and Uttley, P. 2010,  
MNRAS, 404, 931  
\bibitem[Ghez et al.(2008)]{ghez08} Ghez, A.~M., Salim, S., Weinberg, N.~N., Lu, J.~R., Do, T., Dunn, J.~K., Matthews, K., Morris, M.~R., Yelda, S., Becklin, E.~E. and Kremenek, T. et al. 2008, \apj, 689, 1044
\bibitem[Gillessen et al.(2009)]{gillessen09} Gillessen, S. et al. 2009, \apj, 692, 1075
\bibitem[Goldston et al.(2005)]{goldston05} Goldston, J.~E., Quataert, E. and Igumenshchev, I.~V. 2005, \apj, 621, 785
\bibitem[Hawley et al.(2002)]{hawley02} Hawley, J.  F. \& Balbus, S.  A. 2002, \apj, 573, 738
\bibitem[Hughes et al.(1992)]{hughes92} Hughes, P.A., Aller \& Aller, 1992, \apj, 396, 469
\bibitem[Kataoka et al.(2001)]{kataoka01} Kataoka, J., Takahashi, T., Wagner, S.~J. et al. 2001, \apj, 560, 659
\bibitem[Liu et al.(2004)]{liu04} {Liu}, S., {Petrosian}, V., \& {Melia}, F. 2004, \apjl, 611, L101
\bibitem[Macquart and Bower (2006)]{mac06} Macquart, J. P. \& Bower, G.C.  2006, ApJ, 641, 302 
\bibitem[Marrone et al.(2006)]{marrone06} Marrone, D.  P.,  Moran, J.  M.,  Zhao, J.-H. \& Rao, R. 2006, Journal of Physics: Conference Series, Volume 54, Proceedings of "The Universe Under the Microscope - Astrophysics at High Angular Resolution", held 21-25 April 2008, in Bad Honnef, Germany. Editors: Rainer Schoedel, Andreas Eckart, Susanne Pfalzner and Eduardo Ros, pp. 354-362
\bibitem[Melia \& Falcke(2001)]{melia01} Melia, F. \& Falcke, H. 2001, \araa, 39, 309
\bibitem[Maitra et al.(2009)]{maitra09} Maitra, D., Markoff, S. \& Falcke, H. 2009, \aap, 508, L13
\bibitem[Mauerhan et al.(2005)]{mauerhan05} Mauerhan, J. C., Morris, M., Walter, F. \& Baganoff, F. K. 2005, \apj, 623, L25
\bibitem[McHardy et al.(2006)]{mchardy06} McHardy, I.~M., Koerding, E., Knigge, C., Uttley, P. et al. 2006, \apj, 444, 730
\bibitem[Meyer et al.(2008)]{meyer08}Meyer, L., Do, T., Ghez, A., Morris, M. R., et al. 2008, \apj, 688, L17
\bibitem[Meyer et al.(2009)]{meyer09} Meyer, L., Do, T., Ghez, A., Morris, M. R., et al. 2009, \apj, 694, L87
\bibitem[Moscibrodzka et al.(2009)]{moscibrodzka09} Moscibrodzka, M., Gammie, C.~F., Dolence, J.~C. 2009, \apj, 706, 497
\bibitem[Porquet et al.(2008)]{porquet08} Porquet, D.,  Grosso, N., Predehl, P., Hasinger, G., Yusef-Zadeh, F.,  Aschenbach, B., et al.  2008, \aap, 488, 549
\bibitem[Reid \& Brunthaler(2004)]{reid04} Reid, M. J. \& Brunthaler, A. 2004, \apj, 616, 872
\bibitem[Roberts et al.(1987)]{roberts87} Roberts, D. H.,  Lehar, J., and Dreher, J. W. 1987,  ApJ, 93,   968
\bibitem[Simonetti et al.(1985)]{simonetti85} Simonetti, J. H., Cordes, J. M. and Heeschen, D. S. 1985, \apj, 296, 46
\bibitem[Uttley et al.(2002)]{uttley02} Uttley, P., McHardy, I. M., and Papadakis, I. E. 2002,  MNRAS, 332, 231 
\bibitem[van der Laan(1966)]{laan66} van der Laan, H. 1966, \nat, 211, 1131
\bibitem[Yuan et al.(2003)]{yuan03} {Yuan}, F., {Quataert}, E., \& {Narayan}, R. 2003, \apj, 598, 301
\bibitem[Yusef-Zadeh et al.(2006)]{zadeh06} Yusef-Zadeh, F., Roberts, D.,  Wardle, M., Heinke, C.~O. and Bower, G.~C.  2006, \apj, 650, 189
\bibitem[Yusef-Zadeh et al.(2009)]{zadeh09} Yusef-Zadeh, F.,  Bushouse, H.,  Wardle, M.,  Heinke, C. et al.  2009, 
ApJ, 706, 348
\end{thebibliography}
\end{document}